\documentclass[%
 aip,
 amsmath,amssymb,
 reprint,%
]{revtex4-1}

\usepackage{graphicx}
\usepackage{dcolumn}
\usepackage{bm}
\usepackage{braket}
\usepackage{multirow}
\usepackage{comment}
\usepackage{hyperref}
\usepackage{cleveref}
\usepackage{xspace} 
\usepackage{tikz}
\usetikzlibrary{arrows, arrows.meta}
\usepackage[list=true]{subcaption}
\usepackage{multirow}
\usepackage{dsfont}
\usepackage{natbib}

\newcommand{\abs}[1]{\left| {#1} \right|}              
\newcommand{\norma}[1]{\left \| {#1} \right \|}        
\newcommand*{\wfq}{$\omega$FQ\xspace}
\newcommand*{\imm}{\mathop{}\!\mathrm{i}}
\renewcommand{\Im}{\mathrm{Im}}

\usepackage[utf8]{inputenc}
\usepackage[T1]{fontenc}
\usepackage{mathptmx}

\begin{document}

\preprint{AIP/123-QED}

\graphicspath{{img/}} 

\title[]{Fully Atomistic Modeling of Realistic Plasmonic Materials: Assessing the Performance of Iterative Solvers}


\author{Piero Lafiosca}
\affiliation{Scuola Normale Superiore,
             Piazza dei Cavalieri 7, 56126 Pisa, Italy.}
\author{Tommaso Giovannini}
\affiliation{Scuola Normale Superiore,
             Piazza dei Cavalieri 7, 56126 Pisa, Italy.}
\author{Michele Benzi}
\affiliation{Scuola Normale Superiore,
             Piazza dei Cavalieri 7, 56126 Pisa, Italy.}
\author{Chiara Cappelli}
\email{chiara.cappelli@sns.it}
\affiliation{Scuola Normale Superiore,
             Piazza dei Cavalieri 7, 56126 Pisa, Italy.}
\date{\today}

\begin{abstract}
The fully atomistic modeling of real-size plasmonic nanostructures is computationally demanding, therefore most calculations are limited to small-to-medium sized systems. However, plasmonic properties strongly depend on the actual shape and size of the samples. In this paper we substantially extend the applicability of classical, fully atomistic approaches by exploiting state-of-the-art numerical iterative Krylov-based techniques.  
In particular, we focus on the recently developed \wfq model, when specified to carbon nanotubes, graphene-based nanostructures and metal nanoparticles. The performance of Generalized Minimal Residual (GMRES) and Quasi-Minimum Residual (QMR) algorithms is studied, with special emphasis on the dependence of the convergence rate on the dimension of the structures (up to 1 million atoms) and the physical parameters entering the definition of the atomistic approach.
\end{abstract}

\maketitle

\section{Introduction}

Nanoplasmonics is an emerging field that has been significantly developed in the last decade.\cite{liz2014nanoplasmonics,stockman2011nanoplasmonics} The optical properties of electron-free nanomaterials (e.g. metallic nanoparticles or graphene) are characterized by the rise of surface plasmons, which are coherent oscillations of the conduction electrons induced by an incoming radiation.\cite{maier2007plasmonics} One of the peculiarities of such systems is that their plasmon resonance frequency (PRF) can be tuned as a function of the shape, size and supramolecular structure.\cite{chen2008tunable,diez2009color,stewart2008nanostructured,lohse2013quest} 
In the particular case of graphene, PRF can also be tuned by exploiting electrical gating and chemical doping, which result in a modification of its Fermi energy, thus giving rise to many diverse applications, also in the technological field.\cite{neto2009electronic} 

A peculiarity of plasmonic substrate is that their optical response strongly depends on the system's size. In fact, small nanostructures do not experimentally exhibit the properties of larger structures.\cite{cox2016quantum}
Therefore, only the latter are actually exploited in real applications. These features strongly limit the role of modeling through full quantum mechanical (QM) descriptions, that are impracticable for real-size systems. For this reason, plasmonic structures are generally described by resorting to classical approaches\cite{jensen2008electrostatic,jensen2009atomistic,morton2010discrete,morton2011discrete,payton2012discrete,payton2014hybrid,chen2015atomistic,zakomirnyi2019extended,li2014electronic,zakomirnyi2020plasmonic}, and in particular through implicit models that describe the nanostructure as a continuum, and its optical properties in terms of the frequency-dependent complex permittivity.\cite{draine1988discrete,draine1994discrete,draine2013user,sutradhar2008symmetric,hohenester2012mnpbem,hohenester2014simulating,corni2001enhanced} Clearly, continuum methods lack any atomistic description of the substrates, which can instead be retained by using explicit, atomistic approaches, where the optical response directly arises from atomic parameters, as for
instance computed frequency-dependent atomic complex polarizabilities.\cite{jensen2008electrostatic,jensen2009atomistic,morton2010discrete,morton2011discrete,payton2012discrete,payton2014hybrid,chen2015atomistic,giovannini2019classical,giovannini2020graphene} On the other hand, continuum approaches are not computationally demanding, because the Maxwell equations are solved on the \textit{surface} of the nanostructure, while atomistic models require the treatment of all atoms, therefore the cost of the calculation scales as a function of the \textit{volume}. This is the price to pay to describe for instance doping\cite{n2012localized,manjavacas2012plasmon} and structural defects\cite{anand2014role}, which cannot be modelled by continuum approaches. 

Due to the unfavourable scaling of classical atomistic models with the number of atoms, the calculation of the optical properties of real size structures, which are constituted by millions of atoms, is a hard task. Regardless of the specific discrete model, in fact, the calculation of the plasmonic response involves the solution of a number of complex-valued, frequency-dependent linear systems, with a dense coefficient matrix of order equal to the number of atoms. The solution of the linear system is challenging in terms of both memory requirements and computational cost, therefore the investigation of the performance and applicability of alternative numerical methods to approach this problem is particularly relevant and timely. 

In this work we apply different numerical techniques to solve the problem. In particular, we model the optical response of large graphene-based structures by exploiting a fully atomistic, yet classical approach that has been recently developed by some of us, named \wfq.\cite{giovannini2019classical,giovannini2020graphene,bonatti2020plasmonic} There, each atom of the substrate is endowed with a charge, whose value is determined by solving a complex-valued linear system. The charge exchange between the atoms is governed by the Drude mechanism and by quantum tunneling. \wfq has recently been shown to be successfully applicable to both graphene-based materials and plasmonic metal nanoparticles, for which accurate results are obtained. However, similarly to all atomistic approaches, the solution of the \wfq linear system by means of direct methods based on the LU factorization of the system matrix scales unfavourably with the size of the nanostructure, therefore it is an ideal playground to test the performance of various solution algorithms based on Krylov subspace iterative methods,\cite{van2003iterative} such as the Generalized Minimal Residual (GMRES) and Quasi-Minimum Residual (QMR) algorithms.\cite{saad1986gmres,freund1991qmr} In particular, GMRES and QMR computational timing and convergence rates are compared with the direct inversion solution, which is also much more demanding in terms of memory requirements. 

The manuscript is organized as follows. In the next section, we briefly recap the fundamentals of \wfq, by especially focusing on the mathematical properties of the \wfq linear system. Similarities and differences with continuum approaches are discussed, and Krylov-based methods are presented. We then apply the newly developed algorithms to the modeling of the plasmonic properties of selected nanostructures, of size up to hundreds of nanometer and constituted by roughly one million atoms. A summary and an overview of future developments end the manuscript.

\section{Theoretical Model}


\wfq is a fully atomistic, classical model that describes the response of a system, such as nanoparticles or graphene-based nanostructures, to the external electric field $\mathbf{E}$.\cite{giovannini2019classical} In particular, each atom of the system is endowed with a charge. Charge exchange between different atoms is governed by the Drude mechanism of conduction,\cite{jackson2007classical} and modulated by quantum tunneling.\cite{giovannini2019classical} The key equation for solving the charges $\mathbf{q}$ reads:
\begin{equation}\label{eq:wfq-original}
\sum_{j=1}^N\bigg(\sum_{\substack{k=1 \\ k\ne i}}^N K_{ik}^\mathrm{tot} (D_{kj} - D_{ij})+ \imm\omega\bigg)q_j = \sum_{\substack{j=1 \\ j\ne i}}^N(V^\mathrm{ext}_i-V^\mathrm{ext}_j)K^\mathrm{tot}_{ij},
\end{equation}
where $V^\mathrm{ext}_i$ is the electric potential acting on the $i$-th charge associated with the external electric field oscillating at frequency $\omega$, $D_{ij}$ is the charge interaction kernel and $K^\mathrm{tot}_{ij}$ is a matrix accounting for both Drude and tunneling mechanisms.

More in detail, the linear system in \cref{eq:wfq-original} describes the response of a set of $N$ complex-valued charges $q_j$ under the effect of an external monochromatic uniform electric field of frequency $\omega$ polarized along the $\hat{\mathbf{k}}$ direction, with $\hat{\mathbf{k}} = \hat{\mathbf{x}},\hat{\mathbf{y}},\hat{\mathbf{z}}$. The associated potential $V^\mathrm{ext}$ on each atom, entering the right-hand side of \cref{eq:wfq-original}, is defined as:
\begin{equation}\label{eq:potential}
V^\mathrm{ext}_i = V^\mathrm{ext}(\mathbf{r}_i) = - E^k_{0} k_i, \qquad i=1,\dots,N,
\end{equation}
where $E^k_{0}$ is the intensity of the electric field along the $k$ direction, $\mathbf{r}_i$ is the position of the $i$-th charge and $k_i$ is the component of $\mathbf{r}_i$ along the $\hat{\mathbf{k}}$ axis, i.e. $k_i = \mathbf{r}_i\cdot\hat{\mathbf{k}}$. 
The matrix $\mathbf{D}$ on the left-hand side of \cref{eq:wfq-original} describes the electrostatic interaction between the charges, and it is defined in the standard formulation of the FQ force field exploited for treating molecular systems.\cite{giovannini2020molecular,giovannini2020theory} In order to avoid the so-called ``polarization catastrophe'', \cite{jensen2008electrostatic} instead of point charge we use spherical Gaussian charge distributions of width $d_i$ and $d_j$ to describe \wfq charges. $\mathbf{D}$ elements then read:\cite{lipparini2011polarizable,giovannini2019classical,giovannini2019polarizable,mayer2007formulation}
\begin{equation}\label{eq:d-matrix}
D_{ij} = 
\begin{cases}
\frac{1}{r_{ij}}\mathrm{erf}\bigg(\frac{r_{ij}}{\sqrt{d_i^2+d_j^2}}\bigg) & i \ne j \\
\eta_i & i = j
\end{cases},\quad i,j=1,\dots,N,
\end{equation}
where $r_{ij} = \abs{\mathbf{r}_i-\mathbf{r}_j}$ is the distance between charges $i$-th and $j$-th, $\mathrm{erf}$ is the error function and $\eta_i$ is the atomic chemical hardness of the $i$-th atom.\cite{lipparini2011polarizable,rick1994dynamical} Gaussian widths $d_i$ and $d_j$ are chosen for each atom by imposing that the limit for $\mathbf{r}_i\rightarrow\mathbf{r}_j$ corresponds to the diagonal element of the matrix, i.e.:
\begin{equation}
\lim_{\mathbf{r}_i\rightarrow\mathbf{r}_j} \frac{1}{r_{ij}}\mathrm{erf}\Bigg(\frac{r_{ij}}{\sqrt{d_i^2+d_j^2}}\Bigg) = \eta_j\quad\Rightarrow\quad d_i = \sqrt{\frac{2}{\pi}}\frac{1}{\eta_j}.
\end{equation}
The \textbf{D} matrix can be formally seen as an overlap matrix defined in the scalar product weighted by the $\frac{1}{r}$ function. Therefore, it is symmetric positive definite (SPD).\cite{lipparini2011polarizable} 
The $\mathbf{K}^\mathrm{tot}$ matrix in \cref{eq:wfq-original} reads:
\begin{equation}\label{eq:k-tot-matrix}
\begin{split}
K^\mathrm{tot}_{ij} = & (1-f(r_{ij}))K^\mathrm{dru}_{ij}(\omega) \\
= &(1-f(r_{ij})) \frac{2n_0}{1/\tau-\imm\omega}\frac{\mathcal{A}_{ij}}{r_{ij}}, \quad i,j=1,\dots,N,\ \ i\ne j,
\end{split}
\end{equation}
where $n_0$ is the electron density, $\tau$ is a friction-like constant due to scattering events and $\mathcal{A}_{ij}$ is an effective area dividing atoms $i$ and $j$. $f$ is a Fermi-like damping function, defined as:
\begin{equation}\label{eq:fermi}
f(r_{ij}) = \frac{1}{1 + \exp\bigg[-d\bigg(\frac{r_{ij}}{s\cdot r_{ij}^0}-1\bigg)\bigg]},
\end{equation}
in which $r_{ij}^0$ is the equilibrium distance between atoms $i$ and $j$, while $d$ and $s$ are parameters ruling the shape of the damping function. $\mathbf{K}^\mathrm{tot}$ is a frequency-dependent symmetric complex-valued matrix, and it can be interpreted as a ``dynamic'' response matrix, whereas the \textbf{D} matrix describes the ``static'' response. It is worth noticing the expression of $\mathbf{K}^\mathrm{tot}$ can be associated with two alternative response regimes. When $f(r_{ij})$ goes to zero, the purely Drude conductive regime is recovered; as $r_{ij}$ increases, the electron transfer descreases exponentially, thus leading to the typical tunnelling mechanism.\cite{giovannini2019classical} The diagonal elements of $\mathbf{K}^\mathrm{tot}$ do not enter \cref{eq:wfq-original}, but the notation can be simplified by imposing $K^\mathrm{tot}_{ii} = 0$ for all $i=1,\dots,N$ and extending the summations over $k$ and $j$ in \cref{eq:wfq-original} to all $N$ atoms of the system.

As a final remark, the electron density $n_0$, that appears in \cref{eq:k-tot-matrix}, is a specific property of the chemical composition of the plasmonic substrate and of the shape of the system. In a general 3D system, $n_0$ can be expressed as $n_0 = \frac{\sigma_0/\tau}{m^\star}$, where $\sigma_0$ is the static conductance of the material, while $m^\star$ is its effective electron mass, which can be approximated to 1 for metal nanoparticles. However, in case of graphene-based  materials, such as graphene sheets or carbon nanotubes, the effective electron mass needs to be taken into account.\cite{giovannini2020graphene}
In graphene-based materials, $m^\star$ can be expressed as 
\begin{equation}\label{eq:2d-eff-mass}
m^\star = \frac{\sqrt{\pi n_{\mathrm{2D}}}}{v_F},
\end{equation}
where $n_\mathrm{2D}$ is the 2D electron density of the system and $v_F$ is the Fermi velocity.\cite{neto2009electronic} The latter is related to the Fermi energy $\varepsilon_F$ through the expression $v_F = \sqrt{\frac{2\varepsilon_F}{m_0}}$, where $m_0$ is the electron rest mass.\cite{neto2009electronic} The 2D electron density $n_{\mathrm{2D}}$ can be calculated from $n_0$ as $n_\mathrm{2D} = n_0\cdot a_0$, with $a_0$ being the Bohr radius.\cite{giovannini2020graphene} Then, the 2D electron density can be calculated as the ratio of the number of atoms $N$ and the surface of the system $S$, i.e.:
\begin{equation}\label{eq:n-2d}
n_\mathrm{2D} = \frac{\alpha N}{S},
\end{equation}
where $\alpha$ is a parameter ($< 1$) which selects the fraction of $\pi$ electrons which are involved in the studied plasmonic excitation.\cite{giovannini2020graphene} We note that such a parameter is uniquely determined by the value of $\varepsilon_F$.\cite{giovannini2020graphene}

\subsection{Properties of the \texorpdfstring{\wfq}{wFQ} linear system}\label{subsec:properties}

In this section, we analyze the mathematical properties of the \wfq linear system defined in \cref{eq:wfq-original}. 
We first notice that in \cref{eq:k-tot-matrix} the complex frequency-dependent ratio describing the Drude model can be gathered:  
\begin{equation}\label{eq:k-tot-factor}
\begin{split}
K^\mathrm{tot}_{ij} & = w(\omega) \overline{K}^\mathrm{tot}_{ij} = w(\omega) (1-f(r_{ij}))\frac{\mathcal{A}_{ij}}{r_{ij}}, \\ w(\omega) & = \frac{2n_0}{1/\tau-\imm\omega},
\end{split}
\end{equation}
where $\overline{\mathbf{K}}^\mathrm{tot}$ is a symmetric real-valued matrix. The frequency-dependent complex factor $w(\omega)$ is always non-zero, so we can take it out from \cref{eq:wfq-original}:
\begin{equation}\label{eq:manipulation}
\sum_{j=1}^N\bigg(\sum_{k=1}^N \overline{K}_{ik}^\mathrm{tot}(D_{kj}-D_{ij}) + \imm\frac{\omega}{w(\omega)}\bigg)q_j = \sum_{j=1}^N(V^\mathrm{ext}_i-V^\mathrm{ext}_j)\overline{K}^\mathrm{tot}_{ij}.
\end{equation}
At this point we can introduce the following notation:
\begin{align}
A_{ij} & = \sum_{k=1}^N \overline{K}_{ik}^\mathrm{tot}(D_{kj}-D_{ij}) \label{eq:a-matrix-def}\\
z(\omega) & = -\imm\frac{\omega}{w(\omega)} = -\frac{\omega}{2n_0\tau}(\omega\tau+\imm) \label{eq:z-def} \\
R_i & = \sum_{j=1}^N(V^\mathrm{ext}_i-V^\mathrm{ext}_j)\overline{K}^\mathrm{tot}_{ij} \label{eq:r-vector-def},
\end{align}
and \cref{eq:manipulation} can be expressed in vector notation as
\begin{equation}\label{eq:wfq-a-matrix}
(\mathbf{A}-z(\omega)\mathbf{I})\mathbf{q} = \mathbf{R},
\end{equation}
where $\mathbf{I}$ is the $N$-dimensional identity matrix.
It can be noted that \cref{eq:wfq-a-matrix} is fully equivalent to \cref{eq:wfq-original}, but $\mathbf{A}$ this time is a real-valued frequency-independent nonsymmetric matrix, of which the diagonal elements are shifted by a complex quantity. 
Moreover, the $\mathbf{A}$ matrix defined in \cref{eq:a-matrix-def} can be rewritten as:
\begin{equation}\label{eq:a-prodotto}
\begin{split}
A_{ij} & = \sum_{k=1}^N\overline{K}^\mathrm{tot}_{ik}(D_{kj}-D_{ij}) \\
& = \big(\overline{\mathbf{K}}^\mathrm{tot}\mathbf{D}\big)_{ij} -\bigg(\sum_{k=1}^N\overline{K}^\mathrm{tot}_{ik}\bigg)D_{ij}.
\end{split}
\end{equation}
By introducing a diagonal matrix $P_{il} = \big(\sum_{k=1}^N\overline{K}^\mathrm{tot}_{ik}\big)\delta_{il}$ where $\delta_{il}$ is the Kronecker delta, we can write:
\begin{equation}
\bigg(\sum_{k=1}^N\overline{K}^\mathrm{tot}_{ik}\bigg)D_{ij} = \sum_{l=1}^N P_{il}D_{lj},
\end{equation}
and plugging the definition into \cref{eq:a-prodotto} we obtain:
\begin{equation}\label{eq:a-matrix-product}
\mathbf{A} = (\overline{\mathbf{K}}^\mathrm{tot}-\mathbf{P})\mathbf{D}.
\end{equation}
Therefore, the $\mathbf{A}$ matrix can be formulated as the product of two real-valued symmetric matrices, since $\overline{\mathbf{K}}^\mathrm{tot}$ and $\mathbf{D}$ are symmetric and $\mathbf{P}$ is diagonal. However, $\mathbf{A}$ is a nonsymmetric matrix because $\mathbf{D}$ and $\overline{\mathbf{K}}^\mathrm{tot}-\mathbf{P}$ in general do not commute. Nevertheless, the following equality holds:
\begin{equation}\label{eq:a-matrix-d}
\mathbf{DA} = \mathbf{D}(\overline{\mathbf{K}}^\mathrm{tot}-\mathbf{P})\mathbf{D} = \big[(\overline{\mathbf{K}}^\mathrm{tot}-\mathbf{P})\mathbf{D}\big]^T\mathbf{D} = \mathbf{A}^T\mathbf{D},
\end{equation}
where $^T$ indicates the transposition operator. Recalling that $\mathbf{D}$ is an SPD matrix,\cite{lipparini2011polarizable} we can define the $\mathbf{D}$-inner product as:
\begin{equation}\label{eq:d-inner-product}
\forall\ \mathbf{x},\mathbf{y} \in \mathds{C}^N\quad \braket{\mathbf{x},\mathbf{y}}_\mathbf{D} = \braket{\mathbf{Dx},\mathbf{y}},
\end{equation}
where $\braket{\cdot,\cdot}$ is the standard Euclidean inner product. From \cref{eq:a-matrix-d} and \cref{eq:d-inner-product} it can be demostrated that $\mathbf{A}$ is self-adjoint with respect to the $\mathbf{D}$-inner product, i.e.:
\begin{equation}\label{eq:a-d-symmetric}
\braket{\mathbf{Ax},\mathbf{y}}_\mathbf{D} = \braket{\mathbf{DAx},\mathbf{y}} = \braket{\mathbf{A}^T\mathbf{Dx},\mathbf{y}} = \braket{\mathbf{Dx},\mathbf{Ay}} = \braket{\mathbf{x},\mathbf{Ay}}_\mathbf{D}.
\end{equation}
\Cref{eq:a-d-symmetric} allows us to conclude that even if $\mathbf{A}$ is a nonsymmetric matrix, it is self-adjoint with respect to the inner product induced by the SPD matrix $\mathbf{D}$, therefore $\mathbf{A}$ is diagonalizable with real eigenvalues.

As a final remark, the alternative expression of the $\mathbf{A}$ matrix in \cref{eq:a-matrix-product} allows us to derive another property of the matrix itself. 
%
%
In fact, $\overline{\mathbf{K}}^\mathrm{tot}-\mathbf{P}$ is such that each row (or column) sums up to zero, therefore it is a singular matrix. By this, the matrix $\mathbf{A}$ is also singular.
Nevertheless, the existence and uniqueness of the linear system solution in \cref{eq:wfq-a-matrix} is guaranteed through the diagonal shift of the coefficient matrix with the complex scalar $z(\omega)$ defined in \cref{eq:z-def}, which is non-zero when $\omega\ne 0$. However, numerical instabilities in the solution of the linear system can arise when $\omega$ approaches zero because $\mathbf{A}$ is close to singular. 

\subsection{Comparison with continuum approaches}

The atomistic nature of \wfq emerges from all the variables that enter \cref{eq:wfq-original}: charge positions, chemical hardnesses in \cref{eq:d-matrix}, effective areas in \cref{eq:k-tot-matrix} and equilibrium distances in \cref{eq:fermi}, as well as electronic features of the material that enter the Drude model. Such an approach allows us to describe the (macroscopic) plasmonic response of the system in terms of (microscopic) atomistic quantities, regardless of the shape of the system. Therefore, complex effects associated with surface roughness and edge effects are automatically considered. 

As stated in the Introduction, the plasmonic response of complex systems can be described by resorting to continuum approaches, such as the Boundary Element Method (BEM).\cite{de2002retarded,hohenester2012mnpbem} There, the material is treated as a continuum and its electronic properties are synthesized by its frequency-dependent dielectric permittivity function $\varepsilon(\omega)$. The plasmonic response arises as a surface charge density $\sigma(\mathbf{r})$, which is computed by solving Maxwell equations, \textit{via} a reformulation as a boundary integral equation on the material surface. From the computational point of view, the latter is discretized in $N$ surface elements centered at positions $\mathbf{r}_i$, with $i=1,\dots,N$. At the same time, the surface charge density $\sigma(\mathbf{r})$ is discretized in terms of $N$ electric charges. The equation for solving the charges in BEM reads:\cite{hohenester2012mnpbem}:
\begin{equation}\label{eq:bem}
\bigg(2\pi\frac{\varepsilon_\mathrm{out}(\omega)+\varepsilon_\mathrm{in}(\omega)}{\varepsilon_\mathrm{out}(\omega)-\varepsilon_\mathrm{in}(\omega)}\mathbf{I}+\mathbf{F}\bigg)\bm{\sigma} = - \overline{\boldsymbol{\phi}},
\end{equation}
where $\varepsilon_\mathrm{out},\varepsilon_\mathrm{in}$ are the frequency-dependent complex-valued dielectric permittivity functions of the outer (usually vacuo) and inner (the material) region, respectively, $\sigma_i = \sigma(\mathbf{r}_i)$ is the electric charge evaluated on the surface element at position $\mathbf{r}_i$, while $\overline{\phi}_i = \frac{\partial \phi(\mathbf{r}_i)}{\partial n}$ is the surface derivative of the external potential $\phi$ at position $\mathbf{r}_i$. Moreover, $\mathbf{I}$ is the $N$-dimensional identity matrix and $F_{ij}$ is the normal derivative of the Green function, i.e.:
\begin{equation}\label{eq:f-bem}
F_{ij} = F(\mathbf{r}_i,\mathbf{r}_j) = \hat{\mathbf{n}}_i \cdot \nabla_{\mathbf{r}_j} \frac{1}{\abs{\mathbf{r}_i-\mathbf{r}_j}},
\end{equation}
where $\hat{\mathbf{n}}_i$ is the normal vector to the surface at point $\mathbf{r}_i$.

\wfq and BEM are genuinely different in terms of performance and versatility: surface roughness can easily be taken into account through an atomistic approach, while the continuum model needs specific treatments such as perturbative expansions.\cite{trugler2011influence} 
Nevertheless, from the purely algebraic point of view, there are some similarities. First of all, \cref{eq:wfq-a-matrix,eq:bem} have the same structure. In both cases, charges are obtained by solving a dense system of linear equations, in which the left-hand side is written in terms of a nonsymmetric frequency-independent real-valued matrix [$\mathbf{A}$ for \wfq (see \cref{eq:a-matrix-def} and $\mathbf{F}$ for BEM (see \cref{eq:f-bem})] with a uniform complex-valued diagonal shift, which describes the electronic properties of the material at a specific frequency ($z(\omega)$ defined in \cref{eq:z-def} for \wfq and the permittivity in \cref{eq:bem} for BEM). Moreover, it has been shown that the $\mathbf{F}$ matrix in \cref{eq:f-bem} is singular and diagonalizable with real eigenvalues\cite{hohenester2012mnpbem,fuchs1975theory,mayergoyz2005electrostatic,fredkin2003resonant}, similarly to the $\mathbf{A}$ matrix in \cref{eq:a-matrix-def} (see \cref{subsec:properties} for the proof). Therefore, the same computational techniques to solve the linear system, which in this paper are described for the \wfq approach, can also be exploited for BEM.

\section{Solution strategies}\label{sec:solution}

In order to model the optical spectra of plasmonic substrates, \cref{eq:wfq-original}, or equivalently \cref{eq:wfq-a-matrix}, need to be solved for a certain number of frequencies $\omega$. This can effectively achieved only by resorting to efficient methods to solve the dense nonsymmetric complex linear system. This point is crucial, especially when the dimensionality of the system increases. Direct techniques of solution (e.g. based on factorization of the coefficient matrix) are to be avoided, because they are inefficient both in terms of storage demand and computational cost, which scales as $N^3$. Thus, matrix-free iterative techniques, which in our implementation scale as $N^2$, are a promising alternative. In particular, such approaches can be implemented without necessarily storing in memory the whole matrix $\mathbf{A}-z(\omega)\mathbf{I}$, but in terms of matrix-vector products which can efficiently be computed in a parallel environment.

One of the most powerful techniques to solve a linear system $\mathbf{Ax}=\mathbf{b}$ (with a generic matrix $\mathbf{A}$) of order $N$ is to resort to Krylov subspace iterative methods.\cite{saad2003iterative,meurant2020krylov} The idea behind this family of methods is to build an approximate solution to the linear system at step $m$ in the $m$-dimensional affine subspace $\mathbf{x}_0 + \mathcal{K}_m(\mathbf{A},\mathbf{r}_0)$, where $\mathcal{K}_m(\mathbf{A},\mathbf{r}_0)$ is the Krylov subspace defined as:
\begin{equation}
\mathcal{K}_m(\mathbf{A},\mathbf{r}_0) = \mathrm{span}\{\mathbf{r}_0,\mathbf{Ar}_0,\dots,\mathbf{A}^{m-1}\mathbf{r}_0\},\quad \mathbf{r}_0 = \mathbf{b}-\mathbf{Ax}_0,
\end{equation}
where $\mathbf{r}_0$ is the residual associated with the initial guess $\mathbf{x}_0$. 

In this work, two different Krylov-based iterative methods have been tested for the solution of \cref{eq:wfq-a-matrix}, namely the Generalized Minimum RESidual algorithm (GMRES)\cite{saad1986gmres} and the Quasi Minimal Residual (QMR) method.\cite{freund1991qmr} 

\textbf{GMRES.} It is a general approach to nonsymmetric linear systems.\cite{saad1986gmres} At each step $m$, an approximate solution of the linear system is obtained as:
\begin{equation}\label{eq:gmres-solution}
\mathbf{x}_m = \mathbf{x}_0 + \mathbf{V}_m \mathbf{y}_m,
\end{equation}
where $\mathbf{V}_m$ is an orthonormal basis of the $m$-dimensional Krylov space $\mathcal{K}_m(\mathbf{A},\mathbf{r}_0)$. The vector $\mathbf{y}_m$ is determined such that the 2-norm of the residual $\mathbf{r}_m = \mathbf{b} - \mathbf{A}\mathbf{x}_m$ is minimal over $\mathcal{K}_m$. The orthonormal basis of $\mathcal{K}_m$ is obtained via the Arnoldi process\cite{arnoldi1951principle}, and the orthogonal projection of $\mathbf{A}$ onto $\mathcal{K}_m$ leads to an upper Hessenberg matrix $\mathbf{H}_m = \mathbf{V}_m^\dagger \mathbf{A}\mathbf{V}_m$.\cite{golub2013matrix} Therefore the least squares problem can be efficiently solved through a QR factorization of $\mathbf{H}_m$.\cite{golub2013matrix} The QR decomposition of $\mathbf{H}_m$ can be updated cheaply on each iteration, but at each step a new vector must be stored, so the memory cost is not constant during the iterative procedure.\cite{golub2013matrix}

In order to reduce the memory required by GMRES, the so-called ``restarted'' GMRES algorithm has been developed, also known as GMRES($k$)\cite{saad1986gmres,joubert1994convergence}. There, the iterative procedure is stopped after $k$ steps, and the GMRES algorithm is restarted by using the last iterative vector $\mathbf{x}_k$ as the new initial guess vector from which the Krylov subspace is built once again. By this, no more than $k$ vectors are stored in memory at the same time, however the algorithm is expected to converge more slowly than standard GMRES .\cite{joubert1994convergence}

\textbf{QMR.} 
Similarly to GMRES, QMR has been developed for solving nonsymmetric linear systems. In this case, the coupled two-term Lanczos algorithm is adopted to generate two Krylov spaces $\mathcal{K}_m(\mathbf{A},\mathbf{p}_0)$ and $\mathcal{K}_m(\mathbf{A}^T,\mathbf{q}_0)$ where $\mathbf{p}_0,\mathbf{q}_0$ are two initial vectors.\cite{freund1994qmr} At each step, an approximate solution to the complex-valued linear system is defined as:
\begin{equation}
\mathbf{x}_m = \mathbf{x}_0 + \mathbf{P}_m\mathbf{y}_m,
\end{equation}
which is similar to what is done by GMRES(see \cref{eq:gmres-solution}). In fact, $\mathbf{P}_m$ is the basis set of the Krylov space $\mathcal{K}_m(\mathbf{A},\mathbf{p}_0)$, while the $\mathbf{y}_m$ vector is associated with an approximate 2-norm of the residual.\cite{freund1994qmr} In the QMR formalism, the residual $\mathbf{r}_m = \mathbf{b}-\mathbf{Ax}_m$ can be written as
\begin{equation}\label{eq:qmr-residual}
\mathbf{r}_m = \mathbf{M}_{m}(\mathbf{f}_m - \mathbf{N}_m\mathbf{y}_m),
\end{equation}
where the matrices $\mathbf{M}_m,\mathbf{N}_m$ and the vector $\mathbf{f}_m$ are obtained through the basis sets previously defined.\cite{freund1994qmr} In QMR, the vector $\mathbf{y}_m$ is chosen so to minimize the quantity in brackets in \cref{eq:qmr-residual}. In other words, $\mathbf{y}_m$ is the solution of the least squares problem $\min_{\mathbf{y}_m}\lVert\mathbf{f}_m-N_m\mathbf{y}_m\rVert$. Such a procedure yields a ``quasi-minimization'' of the residual, therefore it is expected to converge more slowly than GMRES. On the other hand, at each step a fixed number of vectors need to be calculated, thus memory requirements are constant along the whole iterative procedure.

Note that we have considered the QMR algorithm because a simplified version has been proposed by Freund and Nachtigal in case of $\mathbf{J}$-symmetric coefficient matrices, i.e. such that $\mathbf{A}^T\mathbf{J} = \mathbf{JA}$ for a SPD matrix $\mathbf{J}.$\cite{freund1995software} This property, which we have demonstrated for the \wfq matrix in \cref{subsec:properties} (with $\mathbf{J}=\mathbf{D}$), allows us to simplify the Lanczos process by choosing the starting vectors such that $\mathbf{q}_0 = \mathbf{Jp}_0$. In this way, the computational effort to compute the basis sets of the involved Krylov space is reduced, because the calculation of the matrix-vector product $\mathbf{A}^T\mathbf{x}$ is not needed. As a final remark, in this work the Krylov iterative methods have been applied without resorting to preconditioning techniques. Some basic preconditioners (e.g. the diagonal of $\mathbf{A}-z\mathbf{I}$) have been tested, without yielding any improvement in the convergence rate.

\section{Numerical Results}

The GMRES and QMR algorithms for the solution of the complex \wfq linear system in \cref{eq:wfq-a-matrix} have been implemented in a standalone {\small \textsc{FORTRAN95}} code, named nanoFQ, in a parallel environment through the {\small \textsc{OPENMP}} application programming interface (API).\cite{dagum1998openmp} To apply complex GMRES, nanoFQ has been interfaced with a public domain software developed by Frayss\'e and co-workers.\cite{gmres} 
The QMR-from-BiConjugate Gradients (BCG) algorithm without look-ahead for $\mathbf{J}$-symmetric matrices\cite{freund1993transpose,freund1995software} has been implemented from scratch. All calculations have been performed on a Xeon Gold 5120 (56 cores, 2.2 GHz) cluster node equipped with 128 GB RAM, if not stated otherwise.

The performance of GMRES and QMR algorithms has been computationally compared by calculating the number of iterations (NI) required to converge the solution of the linear system to a pre-defined threshold. The 2-norm of the residual vector has been used as a convergence criterion, 
\begin{equation}\label{eq:norm-residual}
\lVert \mathbf{r}_k \rVert_2 = \lVert \mathbf{R} - (\mathbf{A}-z(\omega)\mathbf{I})\mathbf{q}_k \rVert_2 < T,
\end{equation}
where $\mathbf{q}_k$ is the vector generated at the $k$-th iterative step and $T$ is a user-defined threshold. 

\begin{figure}[!h]
\centering
\subcaptionbox[Short Subcaption]{CNT\label{subfig:cnt}}{
\begin{tikzpicture}
    \node[anchor=south west,inner sep=0] (image1) at (0,0) {\includegraphics[width=0.4\columnwidth]{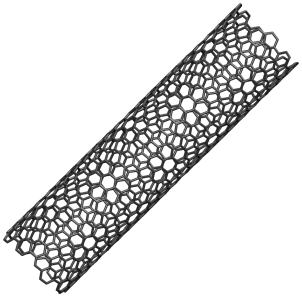}};
    \begin{scope}[x={(image1.south east)},y={(image1.north west)}]
       \draw[latex'-latex',thick] (0,0.35) -- (0.65,1);
       \node[anchor=south east,inner sep=2] at (0.3,0.7) {$L$};
       \draw[latex reversed-latex reversed,thick] (0.25,0.5) -- (0.5,0.25);
       \node[anchor=north west, inner sep=2] at (0.5,0.25) {$d_C$};
       \draw[-latex,thick] (0.8,0.2) -- (0.9,0.3);
       \draw[-latex,thick] (0.8,0.2) -- (0.9,0.1);
       \node[anchor=south west, inner sep=2] at (0.9,0.3) {Z};
       \node[anchor=west, inner sep=2] at (0.9,0.1) {X};
    \end{scope}
\end{tikzpicture}
}
\hfill
\subcaptionbox[Short Subcaption]{GD\label{subfig:gd}}{
\begin{tikzpicture}
    \node[anchor=south west,inner sep=0] (image2) at (0,0) {\includegraphics[width=0.4\columnwidth]{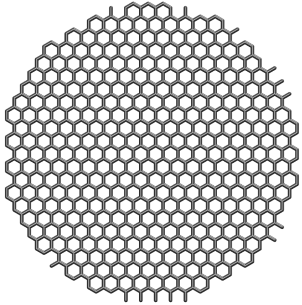}};
    \begin{scope}[x={(image2.south east)},y={(image1.north west)}]
       \draw[latex reversed-latex reversed,thick] (0.12,0.12) -- (0.88,0.88);
       \node[anchor=north east, inner sep=2] at (0.12,0.12) {$d_D$};
    \end{scope}
\end{tikzpicture}
}
\caption{Graphical depiction of CNT (a) and GD (b) molecular structures. The CNT length and diameter ($L$, $d_C$) and the GD diameter ($d_D$) are also highlighted.}
\label{fig:systems}
\end{figure}
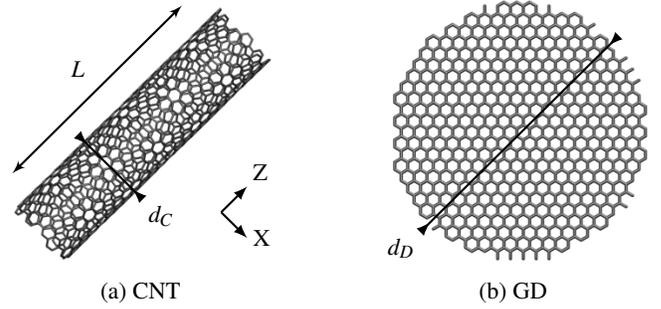

The \wfq approach has been applied to the prediction of the optical properties of selected chiral carbon nanotubes (CNT) and graphene disks (GD) (see \cref{fig:systems} for their molecular structures). For both systems, different geometries have been generated by modifying the length $L$ and/or the diameter $d_C$ for CNTs, and the diameter $d_D$ for GDs. The total number of atoms in the studied structures varies from 8208 to 49248. It is worth remarking that due to the cutting procedure adopted to construct the GDs, dangling bonds possibly occurring on the edges of the disks may be retained (see \cref{subfig:gd}). However, they do not affect the optical properties of large systems (see Fig. S1 given as Supplementary Material - SM). Thus, they can be retained without affecting computed properties and the convergence rate of the two algorithms.

In the following, we study the dependence of NI on:
\begin{itemize}
\item electronic parameters, such as the relaxation time $\tau$ and the Fermi energy $\varepsilon_F$ that enter \cref{eq:k-tot-matrix,eq:2d-eff-mass}, respectively;
\item external field frequency $\omega$, which enters \cref{eq:wfq-a-matrix} through the $z(\omega)$ coefficient defined in \cref{eq:z-def}; 
\item geometry of the systems, in particular the GD diameter and the CNT length/diameter (see \cref{fig:systems});
\item iterative algorithm, i.e. QMR, GMRES or restarted GMRES($k$).
\end{itemize}
Since we are dealing with iterative procedures, also the choice of the initial vector $\mathbf{q}_0$ (see \cref{eq:norm-residual}) can strongly affect the NI. Therefore, in order to compare the different algorithms reliably and in a  reproducible way, we choose $\mathbf{q}_0=0$ in all cases.

\subsection{Electronic parameters}

We first analyze the dependence of NI on the electronic parameters that enter the definition of \wfq model, i.e. the Fermi energy $\varepsilon_F$ (see \cref{eq:2d-eff-mass}) and the relaxation time $\tau$ (see \cref{eq:k-tot-matrix}). 
The plasmonic response of GD varies as a function of both $\tau$ and $\varepsilon_F$.\cite{garcia2014graphene,thongrattanasiri2012quantum,cox2016quantum,giovannini2020graphene}
To analyze such a response, we consider the longitudinal absorption cross section $\sigma_{k}$, that can be calculated as:
\begin{equation}\label{eq:long-cross}
\sigma_k(\omega) = \frac{4\pi\omega}{c}\sum_{i=1}^N \frac{k_i}{E^k_0}\cdot\Im(q_i^k(\omega)),
\end{equation}
where $c$ is the speed of light, $k_i$ is the position of the $i$-th charge along the $\hat{\mathbf{k}}$ axis and $E^k_0$ is the $k$-th component of the intensity of the external electric field. $\Im(q_i^k(\omega))$ is the imaginary part of the $i$-th charge induced by an external electric field polarized along the $\hat{\mathbf{k}}$ axis with frequency $\omega$. The isotropic absorption cross section $\sigma$ can be calculated by averaging the longitudinal one along the three axes, i.e.:
\begin{equation}\label{eq:iso-cross}
\sigma(\omega) = \sum_{k=X,Y,Z} \frac{\sigma_k(\omega)}{3} = \frac{4\pi\omega}{3c}\mathbf{r}_i\cdot\Im(\mathbf{q}(\omega)),
\end{equation}
where we have introduced a compact vector notation in terms of the scalar product between the imaginary part of the charges and their positions. 

It has been shown by some of the present authors\cite{giovannini2020graphene} that PRFs, i.e. frequencies corresponding to $\sigma$ maxima, are independent of $\tau$. In fact, the latter only affects the excitation peak broadening and amplitude, that scale with $\tau$ and $\frac{1}{\tau}$, respectively. 

\begin{figure}[!ht]
\centering
\includegraphics[width=3.4in]{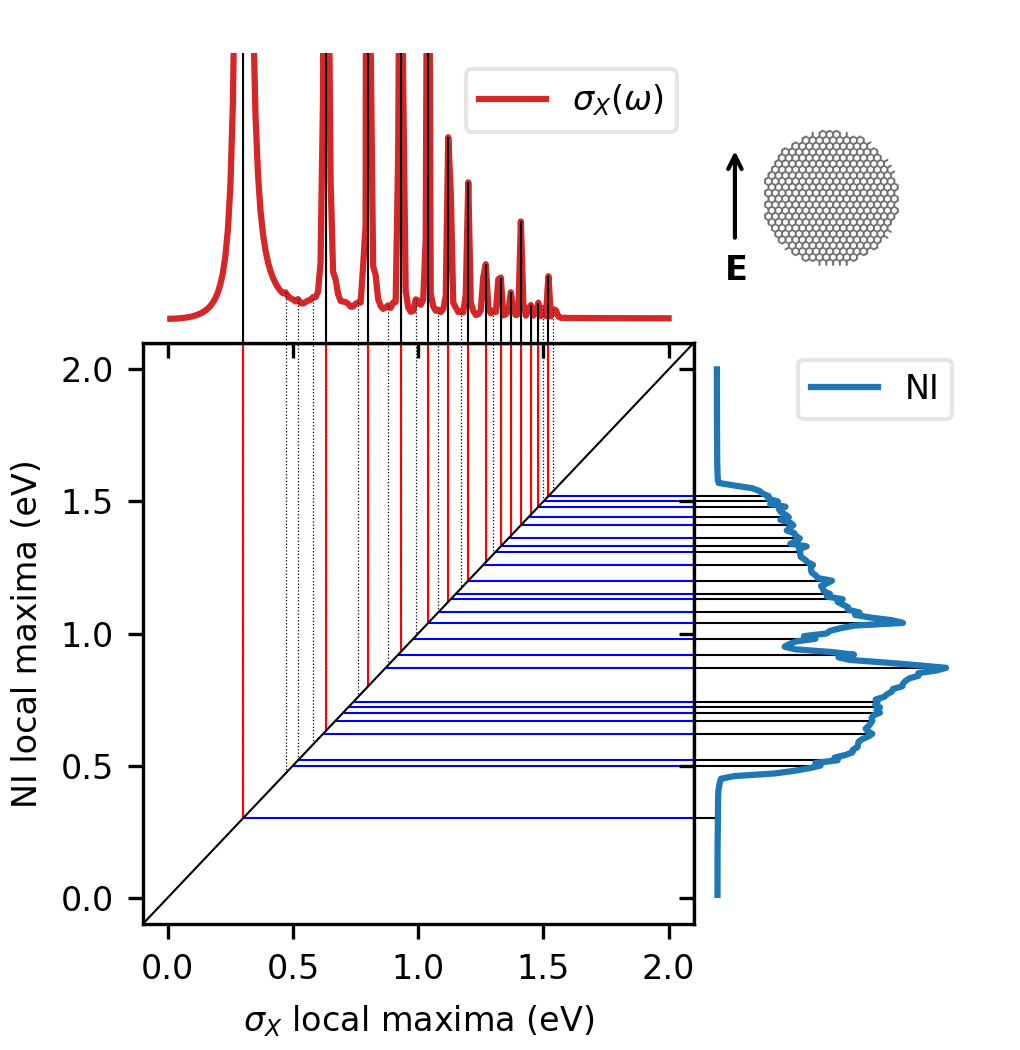}
\caption{Correlation map between GD20 $\sigma_X(\omega)$ (top plot, red line) and NI (right plot, blue line) local maxima.}
\label{fig:corr-disk}
\end{figure}

The dependence of NI on $\tau$ and $\varepsilon_F$ has been studied for a GD with $d_D = 20$ nm (GD20), which is constituted by 11970 carbon atoms. The full (i.e. non-restarted) GMRES NI has been calculated on 200 frequencies in the range between 0.0 eV and 2.0 eV with a constant step of 0.01 eV. The convergence threshold $T$ has been fixed to $10^{-6}$ a.u. (see \cref{eq:norm-residual}). 
\wfq parameters have been set to those exploited in Ref. \citenum{giovannini2020graphene,thongrattanasiri2012quantum} The \wfq linear system has been solved with the GMRES algorithm by using $\varepsilon_F=1.51$ eV and $\tau = 17000$ a.u. The computed $\sigma_X(\omega)$ and NI are reported in \cref{fig:corr-disk}, where the absorption cross section has been scaled to make all peaks visible (see Fig. S2 in the SM).

By varying the external field frequency, $\sigma_X$ shows a set of local maxima of different amplitudes. Since all calculations have been performed with a constant step of 0.01 eV, each PRF has an intrinsic error of 0.01 eV. Similar errors can also affect the relative intensity of the local maxima: the absorption peak is extremely sharp when $\tau$ is large, thus a small variation in frequency induces a large variation in intensity.

\begin{figure}[!ht]
\centering
\includegraphics[width=.48\textwidth]{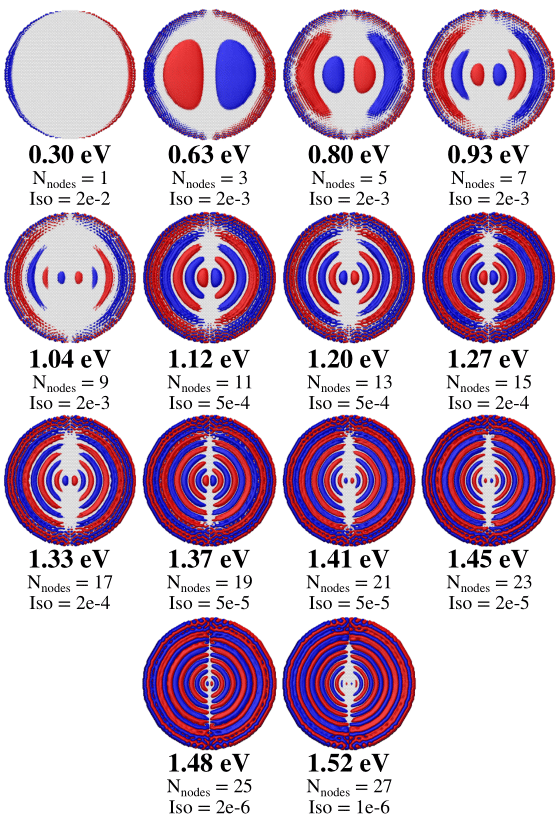}
\caption{Graphical depiction of GD20 plasmon densities calculated at PRFs highlighted in \cref{fig:corr-disk}. The number of nodes ($\text{N}_{\text{nodes}}$) and the isovalue for each plasmon mode are also reported. Densities are obtained by superimposing the Gaussian density associated with each \wfq point charge. \wfq charges have been calculated through the GMRES algorithm by setting $\varepsilon_F=1.51$ eV, $\tau=170$ a.u. and $T=10^{-6}$.}
\label{fig:plasm-170}
\end{figure}

Similarly to $\sigma_X$, the NI plot is characterized by a distribution of local maxima, with the same intrinsic error of PRFs. From an inspection of \cref{fig:corr-disk}, a strong correlation between the two sets of local maximum points is observed, and this is especially true for the most significant maxima highlighted in \cref{fig:corr-disk}. This result is not surprising: from \cref{eq:long-cross} it is expected that a local maximum of $\sigma_X(\omega)$ is necessarily associated with a local maximum (in absolute value) of \wfq point charges. The charge densities associated with $\sigma_X(\omega)$ highlighted local maxima are plotted in \cref{fig:plasm-170}, and they clearly represent plasmon modes of increasing order. In fact, the number of nodes ($\text{N}_{\text{nodes}})$ is always odd for symmetry reason, and increases as frequency increases.\cite{jackson2007classical} Since the iterative procedure starts from the $\mathbf{q}_0 = 0$ vector, when the distance between the guess and the solution vectors increases, NI increases. 

Moving to the global trend of the NI (see \cref{fig:corr-disk}), the required number of iterations is small for the first PRF at $\omega=0.3$ eV, then the NI increases in the middle region of the spectrum and finally decreases. Two underlaying mechanisms may explain this peculiar trend. First, the lowest-order plasmon modes (e.g. the dipolar one at 0.3 eV) are strongly localized on the edge of the system (see \cref{fig:plasm-170}). Therefore a small number of large-valued point charges is involved in the excitation, but most charges are instead close to zero (e.g. those placed in the middle of the structure). By this, the guess vector $\mathbf{q}_0=0$ a.u. is a satisfactory starting point for the iterative procedure, and a small number of iterations is sufficient to obtain the solution vector. This is not true for the highest-order plasmon modes, that are instead delocalized all over the system (see \cref{fig:plasm-170}).
In addition, the plasmonic response intensity (i.e. the point charges absolute value) strongly decreases when the excitation order increases (see \cref{fig:corr-disk}, top), because the number of nodes in the plasmon mode is larger. This is also evinced by the isovalue used to plot the densities in \cref{fig:plasm-170}, which decreases as the PRF increases. Therefore, the NI in correspondence to the highest order plasmon modes tends to decrease.
The presence of low-amplitude local maxima, represented by dashed lines in \cref{fig:corr-disk}, top panel, can be attributed to numerical artifacts (see Fig. S3 in the SM).

\begin{figure}[!ht]
\includegraphics[width=3.4in]{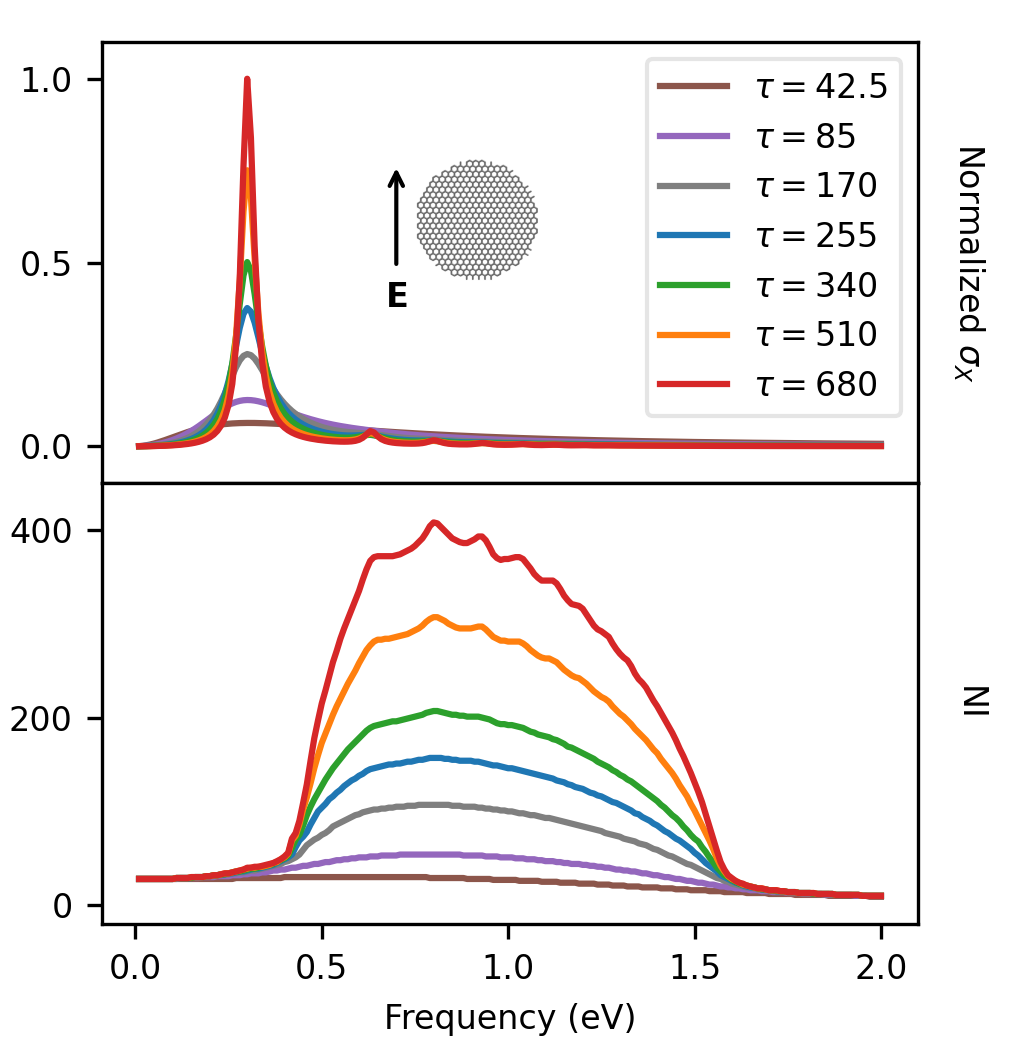}
\caption{GD20 $\sigma_X (\omega)$ (top) and NI (bottom) as a function of $\tau$ (given in a.u.).}
\label{fig:disk-tau}
\end{figure}

\begin{figure}[!ht]
\includegraphics[width=3.4in]{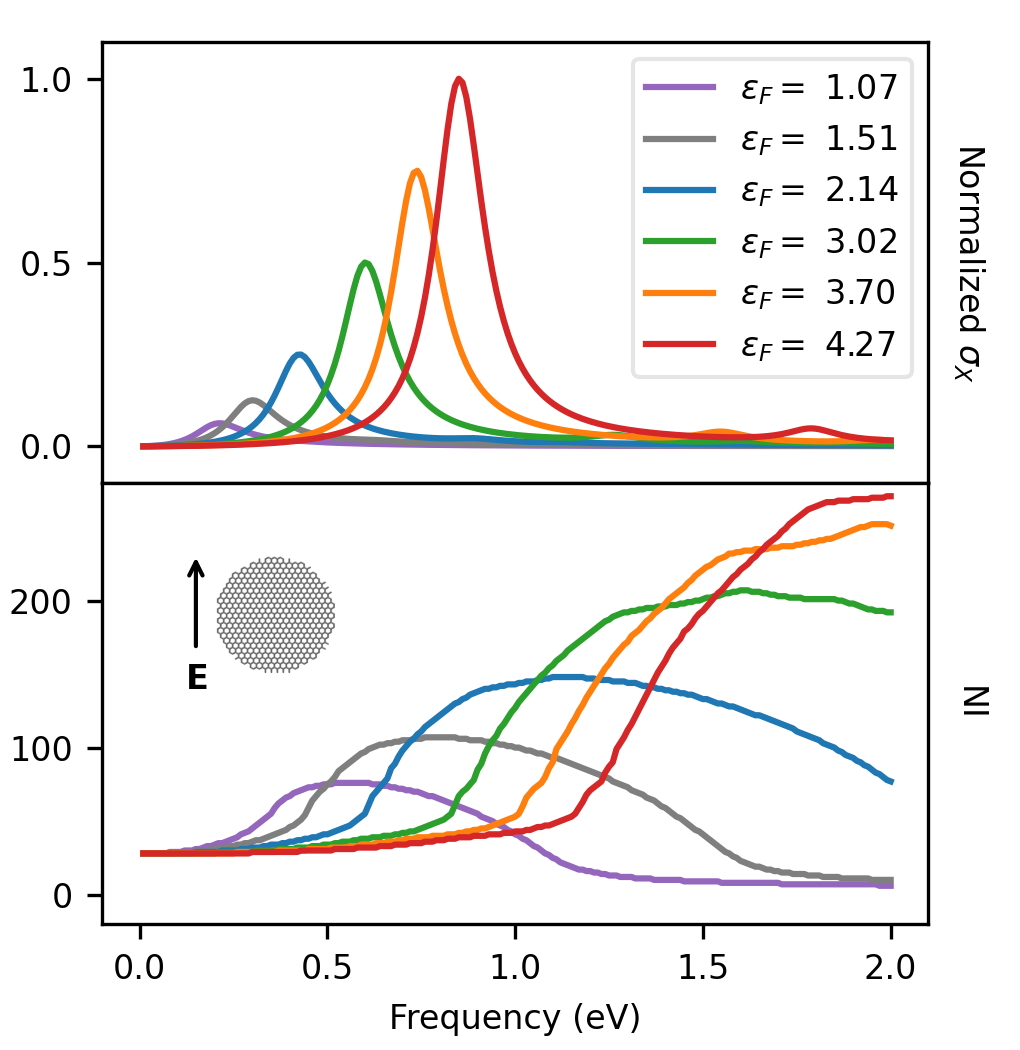}
\caption{GD20 $\sigma_X$ (top) and NI (bottom) as a function of the Fermi energy $\varepsilon_F$ (given in eV).}
\label{fig:disk-ef}
\end{figure}

The dependence of NI and $\sigma_X$ on $\tau$ and $\varepsilon_F$ is reported in \cref{fig:disk-tau,fig:disk-ef}, respectively.
Focusing on the dependence on $\tau$ (\cref{fig:disk-tau}), we see that the PRF is not affected by this parameter, as it has been already reported in previous works.\cite{giovannini2020graphene} The main effect of the variation of $\tau$ is the shrinking of the excitation band shape, and the associated increase of intensity. In the energy region between 0.5 eV and 1.5 eV NI increases with $\tau$, and new local maxima in both $\sigma_X$ and NI at $\tau=680$ a.u. arise. These local maxima are associated to the high-order plasmon resonance modes identified in \cref{fig:corr-disk} and represented in \cref{fig:plasm-170}.

\begin{figure}[!ht]
\includegraphics[width=3.4in]{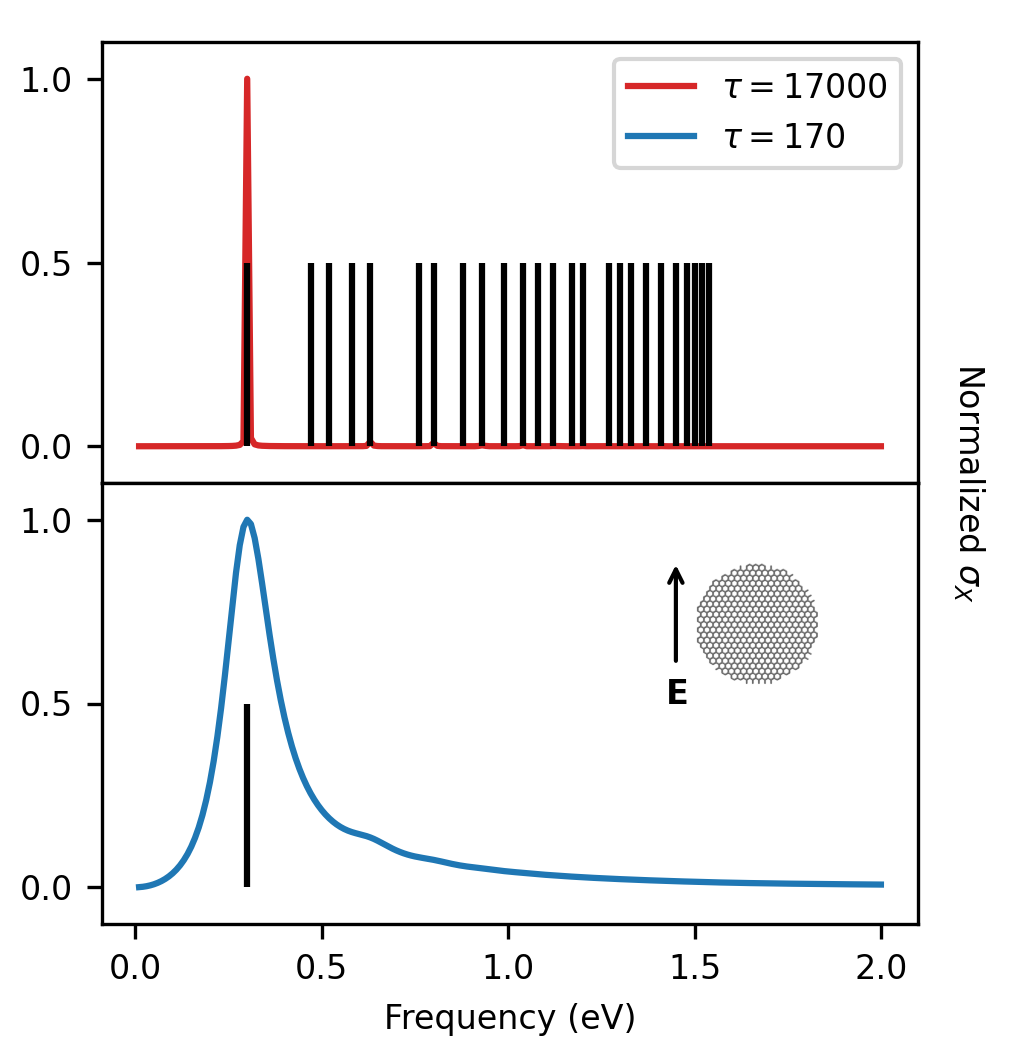}
\caption{GD20 $\sigma_X$ calculated by setting $\tau=17000$ a.u. (top) or $\tau=170$ a.u. (bottom). Local maxima extrapolated from $\sigma_X$ are reported as black sticks.}
\label{fig:disk-17k-170}
\end{figure}

The most relevant plasmon resonance mode is the dipolar excitation, because it is generally associated with the highest amplitude and the lowest PRF, which make it the most suitable for physical applications.\cite{langer2019present} A smaller value of $\tau$ can be adopted to achieve a reliable description of this excitation. In \cref{fig:disk-17k-170}, GD20 $\sigma_X$ calculated by setting $\tau=17000$ a.u. and $\tau=170$ a.u., and $\varepsilon_F=1.51$ eV are reported.
Claerly, the first plasmon excitation with PRF=0.3 eV is the most intense for both $\tau$ values. For $\tau=170$ a.u., $\sigma_X$ is characterized by a single maximum since the reduction of the relaxation time induces also a proportional reduction of the plasmon excitation intensities.\cite{giovannini2020graphene} 
By this, we can conclude that $\tau=170$ a.u. can be chosen in order to reduce the computational cost of the iterative procedure, since we are mostly interested in the description of the dipolar excitation.

The dependence of NI on $\varepsilon_F$ is reported in \cref{fig:disk-ef}. We see that the increase of $\varepsilon_F$ results in a blue shift of the PRF, and in the increase of the absorption intensity. In fact, a higher value of $\varepsilon_F$ is associated with an increase of $n_\mathrm{2D}$ (see \cref{eq:2d-eff-mass}), because a higher fraction of $\pi$ electrons are involved in the excitation. The NI shows a similar trend, because the global maximum is blue shifted and the required number of iterations increases.

\subsection{Geometry}\label{subsec:geometry}

In this section, we investigate the dependence of NI on the geometry of the system. The same CNT and GD structures investigated in the previous section have been selected, for which we have varied the characteristic dimensions (see Fig. \ref{fig:systems}).

\begin{figure}[!h]
\includegraphics[width=3.4in]{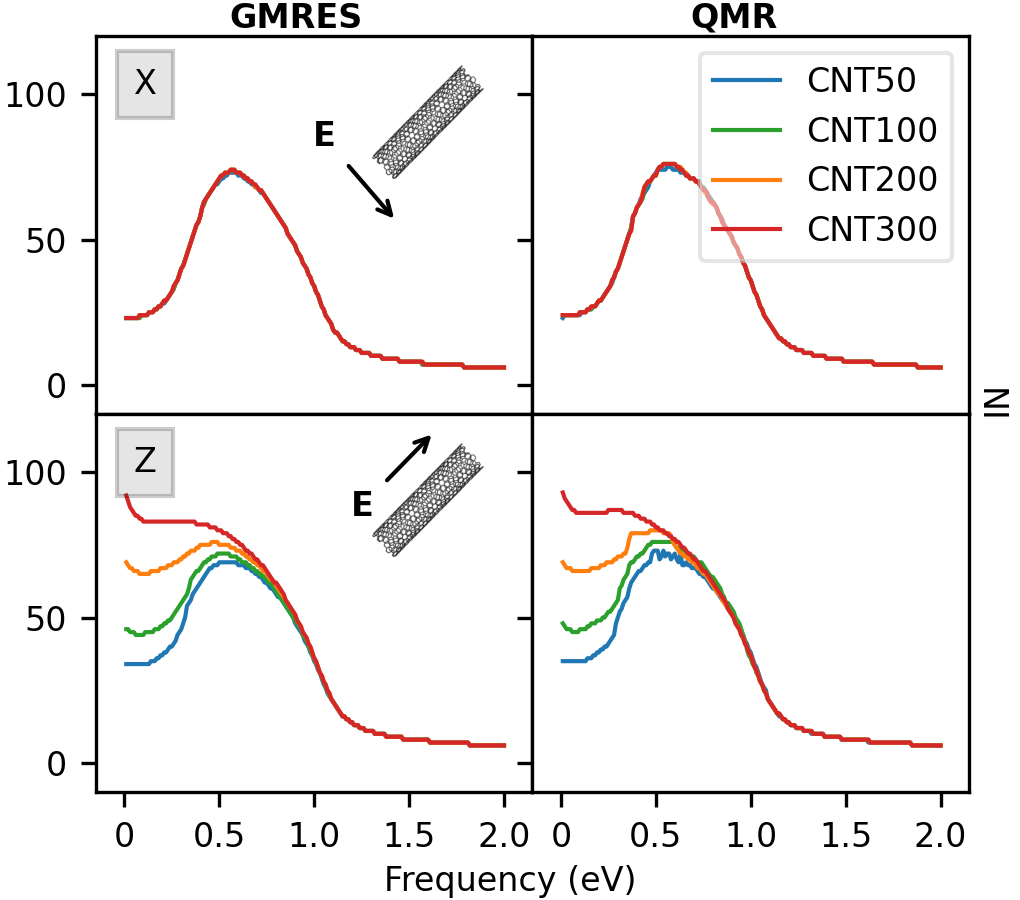}
\caption{Convergence rate dependence on CNT length $L$ (see \cref{tab:cnt-data}) as calculated by using GMRES (left panel) and QMR (right panel). Both longitudinal (bottom) and transversal (top) polarizations of the external field are considered.}
\label{fig:cnt-length-total}
\end{figure}

\textbf{CNT.} \Cref{tab:cnt-data} reports the geometrical parameters of the selected structures.  
\begin{table}[!h]
\begin{center}
\begin{tabular}{lcccc}
\hline
 & $L$ (nm)& Chiral numbers $^a$ & $d_C$ (nm)& Number of C atoms \\
\hline
CNT50 & 50 & \multirow{4}{*}{(8,12)} & \multirow{4}{*}{1.36} & 8208 \\
CNT100 & 100 & & & 16416 \\
CNT200 & 200 & & & 32832 \\
CNT300 & 300 & & & 49248 \\
\hline
CNT1 & \multirow{4}{*}{50} & (8,12) & 1.36 & 8208 \\
CNT2 &  & (16,24) & 2.77 & 16416 \\
CNT3 &  & (24,36) & 4.10 & 24624 \\
CNT4 &  & (32,48) & 5.46 & 32832 \\
\hline
\end{tabular}\\
$^a$ The relation between the diameter $d_C$ and the chiral numbers $(n,m)$ is $d_C = \frac{b_G}{\pi} \sqrt{n^2+m^2+nm}$, where $b_G$ is the graphene lattice basis vector norm, i.e. 0.246 nm\cite{dresselhaus1998physical}
\end{center}
\caption{Geometrical parameters of the studied CNT structures (see \cref{subfig:cnt} for their definition). The number of atoms for each structure is also given.}
\label{tab:cnt-data}
\end{table}
For each structure, the \wfq linear system in \cref{eq:wfq-a-matrix} has been solved for 200 frequencies in the range between 0.0 eV and 2.0 eV with a constant step of 0.01 eV, by setting $\varepsilon_F = 1.04$ eV and $\tau=170$ a.u., respectively.  
First, we comment the results obtained by fixing $d_C = 1.36$ nm and by varying $L$ from 50 to 300 nm (see \Cref{tab:cnt-data}, top block). 
The data are shown in \cref{fig:cnt-length-total} for both GMRES and QMR algorithms, in case the external field is aligned along the transversal (X) or longitudinal (Z) directions. We are then assuming that the two possible transversal directions (X and Y) provide the same polarization, even if all the considered CNTs are chiral. In fact, the differences in the plasmonic response along the two directions are negligible (see Fig. S4 in the SM).
\Cref{fig:cnt-length-total}, top panel, shows that the NI along the transversal direction is the same for all systems, because the diameter is kept constant. In case of longitudinal polarization, the number of iterative steps increases as the length of the system increases in the low energy region of the spectrum, for both GMRES and QMR. This is due to the fact that PRFs are red-shifted as $L$ increases approaching 0 eV (see Tab. S1 in the SM). The $z(\omega)$ factor in \cref{eq:z-def} is therefore close to 0, and since the $\mathbf{A}$ matrix in \cref{eq:wfq-a-matrix} is singular, the number of iterations increases due to increased ill-conditioning.

\begin{figure}[!ht]
\includegraphics[width=3.4in]{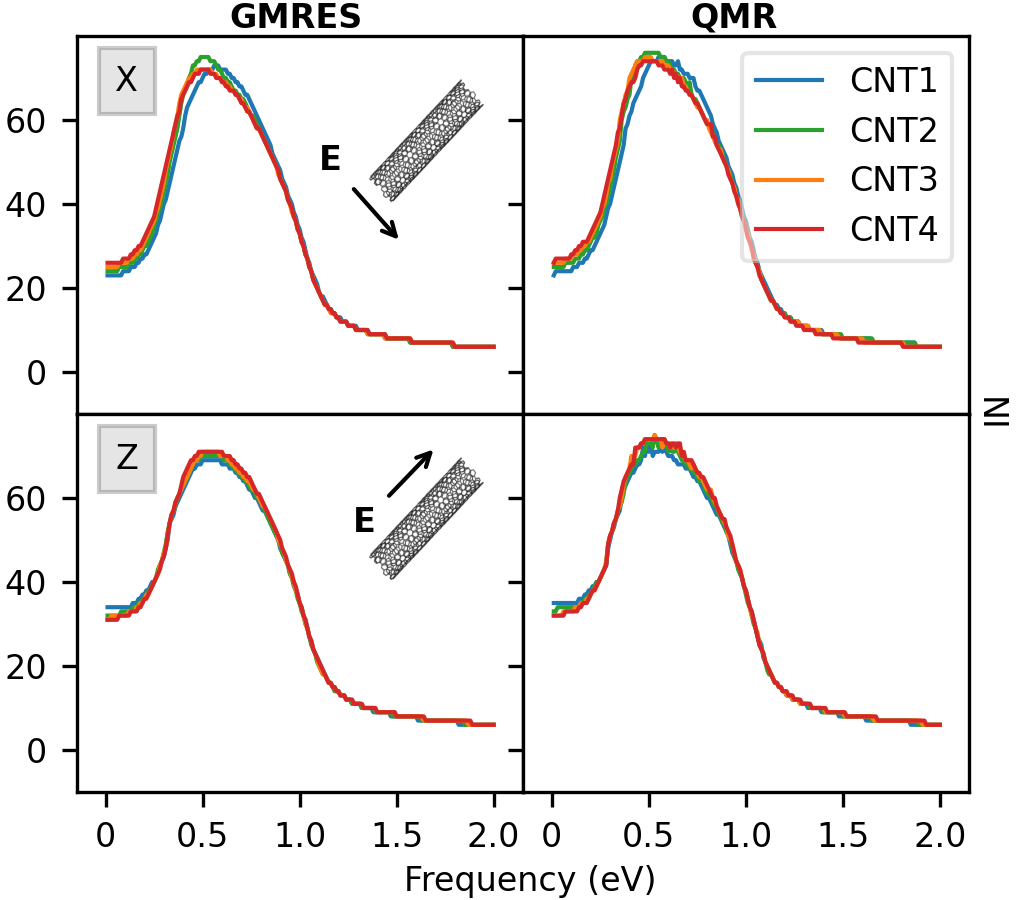}
\caption{Convergence rate dependence on the CNT diameter $d_C$ (see \cref{tab:cnt-data}) as calculated by using GMRES (left panel) and QMR (right panel). Both longitudinal (bottom) and transversal (top) polarizations of the external field are reported.}
\label{fig:cnt-diam-total}
\end{figure}

We now move to comment the results obtained by varying the CNT $d_C$, by keeping constant $L = 50$ nm (see \Cref{tab:cnt-data}, bottom block). Computed GMRES and QMR NI for such systems are reported in \cref{fig:cnt-diam-total}. Differently from the previous case, the NI trend does not show a strong dependence on $d_C$, for both transversal and longitudinal directions of the applied electric field. This is related to the fact that, although PRF energies are red-shifted as $d_C$ increases (along the X direction), the smallest PRF associated with the dipolar plasmon is far from 0 eV (0.37 eV for CNT4, see Tab. S1 in the SM). Therefore, in this case severe ill-conditioning is avoided and the NI remains almost constant.

As a final comment, we note that GMRES and QMR provide almost the same convergence rate. In particular, QMR NI is usually slightly higher than GMRES, thus confirming what has been observed in \cref{sec:solution} (see also Fig. S5 in the SM).

\textbf{GD.} Let us now focus on the NI calculated for four different GDs, obtained by varying the $d_D$ diameter (see \cref{subfig:gd}). Geometrical parameters are reported in \Cref{tab:gd-data}. The \wfq linear systems have been solved by setting the same parameters exploited in case of CNTs, and by imposing $\varepsilon_F = 1.51$ eV. In this case, due to symmetry reason, the external electric field is polarized along one axis only.
\begin{table}[!h]
\begin{center}
\begin{tabular}{lcc}
\hline
ID & $d_D$ (nm)& Number of C atoms \\
\hline
GD20 & 20  & 11970 \\
GD26 & 26  & 20058 \\
GD32 & 32  & 30788 \\
GD36 & 36  & 38974 \\
\hline
\end{tabular}
\end{center}
\caption{Geometrical parameters for the studied GDs (see \cref{subfig:gd} for their definition). The number of atoms for each structure is also given.}
\label{tab:gd-data}
\end{table}

\begin{figure}[!h]
\includegraphics[width=3.4in]{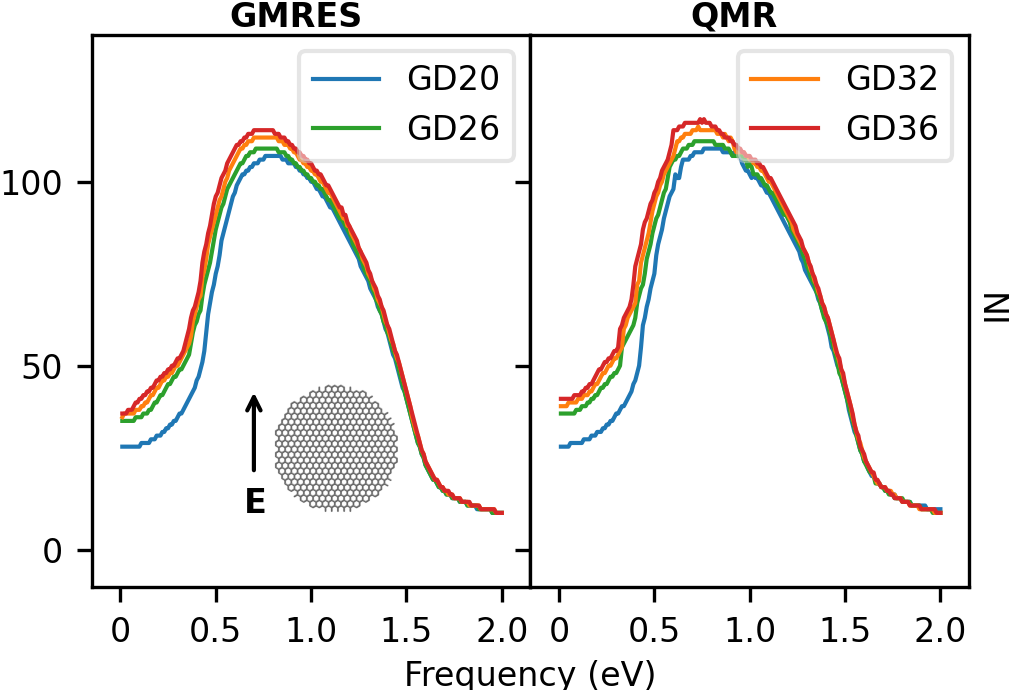}
\caption{Convergence rate dependence on the GD diameter  $d_D$ (see \cref{tab:gd-data}), as calculated by using GMRES (left panel) and QMR (right panel).}
\label{fig:disk-total}
\end{figure}

For each structure, NI has been calculated for GMRES and QMR, and the results are reported in \cref{fig:disk-total}. Interestingly, the NI presents a weak dependence on the $d_D$ diameter. In fact, the largest difference in the number of iterations is of about 10 between GD36 and GD20. However, GD36 has almost four times the atoms of GD20, thus demonstrating the favourable scalability of the two algorithms. We finally note that also in this case the PRFs are red-shifted as the size of the system increases (see Tab. S1 in the SM).

\subsection{Algorithm parameters}

\begin{figure}[!h]
\includegraphics[width=3.4in]{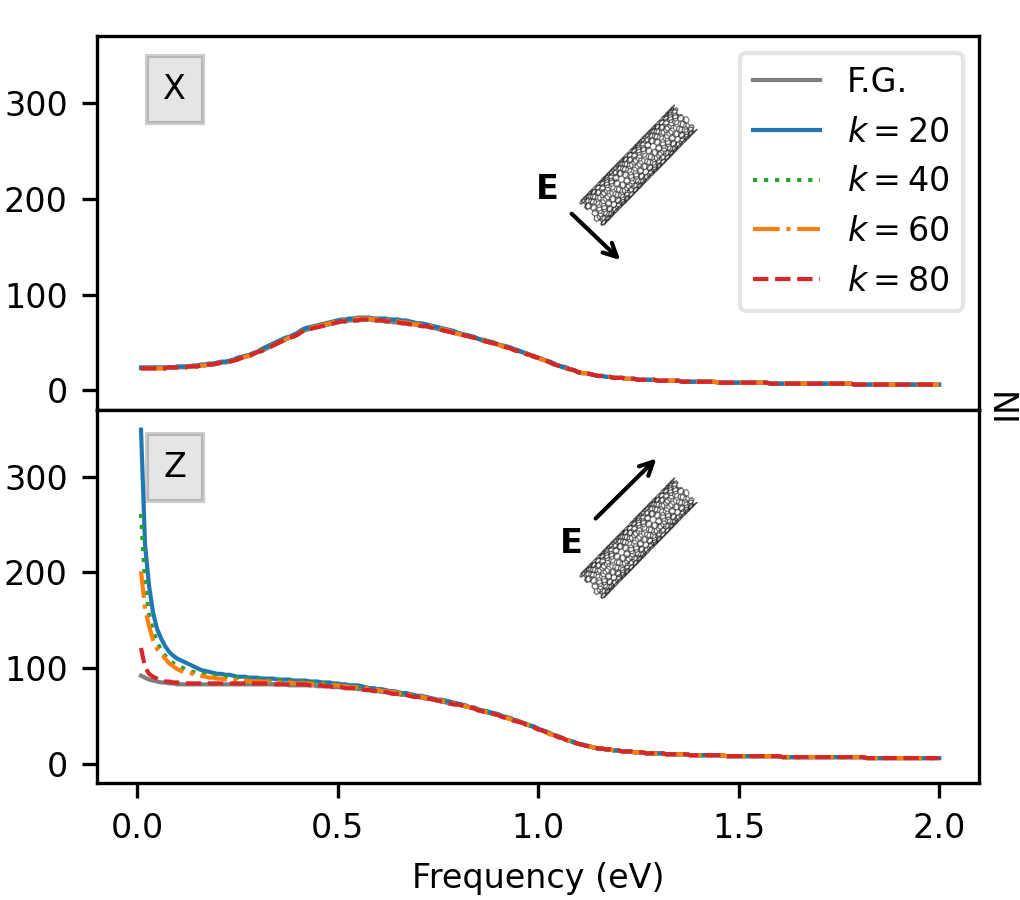}
\caption{CNT300 NI dependence on the dimension of the Krlov subspace $k$ as calculated by using ($k$). The full GMRES (F.G.) values are also shown. Both longitudinal (bottom) and transversal (top) external field is considered.}
\label{fig:restart-300}
\end{figure}

We now move to consider the technical parameters associated with the iterative procedure, and how they affect the convergence rate. We first study the performance of GMRES($k$) (see \cref{sec:solution}), by taking as a reference system the CNT300 structure (see \cref{tab:gd-data}) and exploiting the the same parameters used above for CNTs for solving the \wfq linear system. The iterative procedure has been performed by varying $k$ (between 20 and 80), and by keeping the threshold fixed to $T=10^{-6}$. Computed NI are reported in \cref{fig:restart-300}, together with the corresponding results obtained with the full GMRES (F.G) algorithm (i.e. non-restarted). For X and Y polarizations, the reduction of the Krylov subspace does not affect the NI behaviour. On the contrary, for Z polarization GMRES($k$) and F.G. procedures yield different NI trends in the region between 0 and 0.2 eV. In fact, GMRES($k$) requires a larger number of iterations than F.G. version to reach convergence, independently of the dimension of the Krylov subspace $k$. This is once again due to singularities arising when PRF approaches to 0 eV, which yield ill-conditioning that is exacerbated in the restarted version of the algorithm. Such an explanation is corroborated by the evidence that the number of iterations required to reach convergence decreases by increasing the dimension of the Krylov subspace $k$.
For each $k$, we also notice that GMRES($k$) is almost as efficient as the F.G. procedure in the remaining part of the spectrum. Therefore, in case the PRF is far from 0 eV, the same results can be obtained by using a cheaper iterative procedure in terms of memory requirements, because a smaller number of Krylov basis vectors have to be stored to build up the solution vector. 

\begin{figure*}
\includegraphics[width=6in]{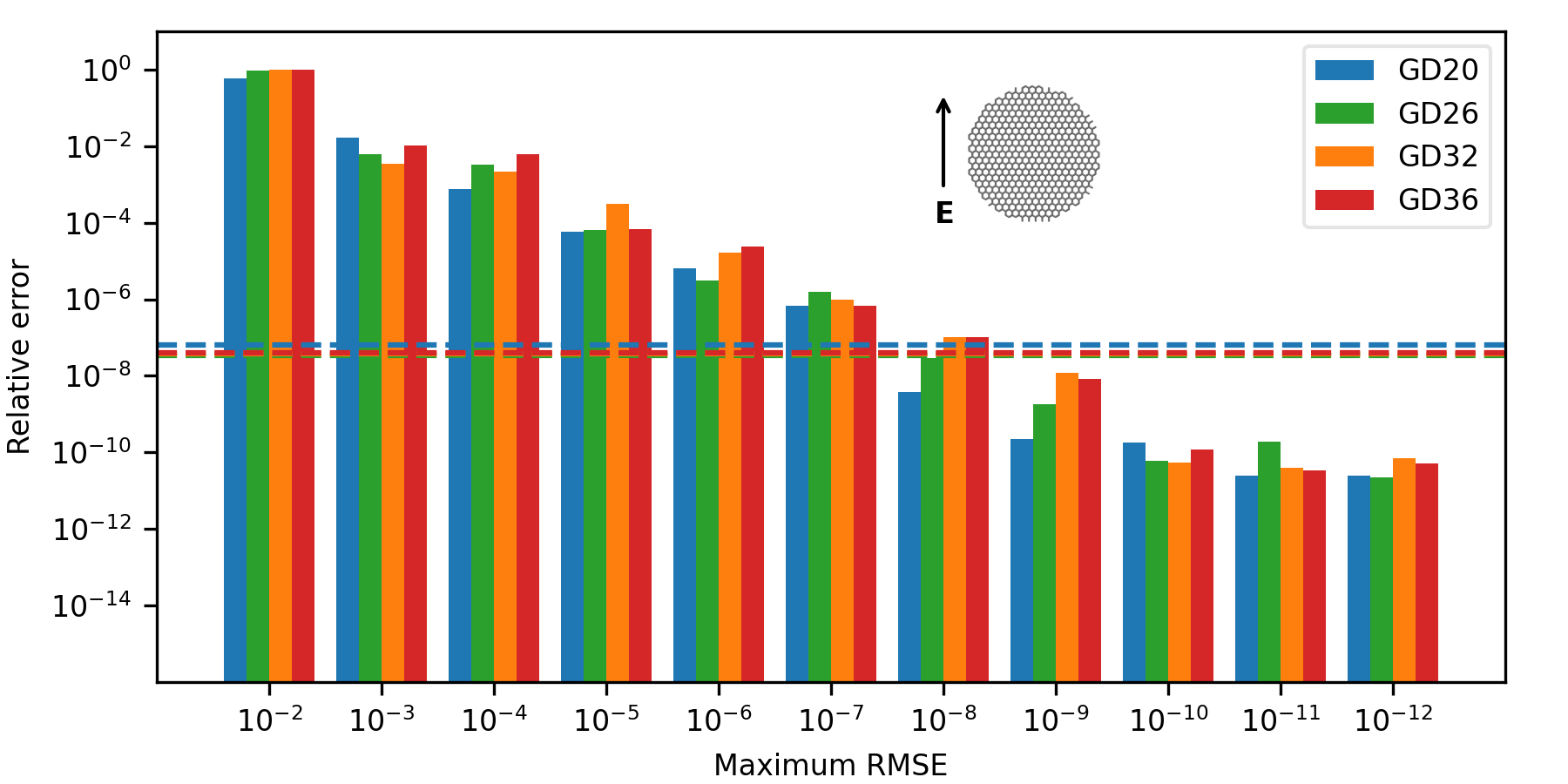}
\caption{Average relative errors for GD20, GD26, GD32 and GD36 (see \cref{tab:cnt-data}) of the GMRES iterative solution with respect to LU factorization of the coefficient matrix at different choices of the RMSE introduced in \cref{eq:rmse}. Dashed lines indicate an approximate upper bound of the intrinsic precision associated with the inversion algorithm.}
\label{fig:thresh-gd}
\end{figure*}

Another quantity that can strongly affect the NI is the threshold $T$ defined in \cref{eq:norm-residual}, which in all previous calculations has been fixed to $10^{-6}$. We notice however that $T$ is independent of the size of the system. Therefore, we can expect that the mean precision of the iterative solution, averaged for each charge, is higher when the number of atoms increases. In fact, \cref{eq:norm-residual} can be rewritten as:
\begin{equation}\label{eq:residual-expression}
\norma{\mathbf{r}_k}_2 = \sqrt{\sum_{i=1}^N \left[R_i - ((\mathbf{A}-z(\omega)\mathbf{I})\mathbf{q}_k)_i\right]^2} < T.
\end{equation}
If we assume that the absolute error is the same for each point charge, i.e. $R_i - ((\mathbf{A}-z(\omega)\mathbf{I})\mathbf{q}_k)_i = \delta q$, and we plug this approximation in \cref{eq:residual-expression} we obtain:
\begin{equation}
\norma{\mathbf{r}_k}_2 \approx \sqrt{N}\delta q < T\quad\Rightarrow\quad \delta q < \frac{T}{\sqrt{N}},
\end{equation}
Therefore, for a given threshold $T$ the absolute error on each charge decreases when the number of atoms $N$ increases.

In order to obtain a size-independent estimate of the accuracy of the iterative solution over all the \wfq charges, we can introduce the following definition of root mean squared error (RMSE):
\begin{equation}
\mathrm{RMSE}(\mathbf{x}_k) = \frac{\norma{\mathbf{R}-(\mathbf{A}-z(\omega)\mathbf{I})\mathbf{x}_k}_2}{\sqrt{N}}.
\end{equation}
Then, we can define a new convergence criterion as:
\begin{equation}\label{eq:rmse}
\begin{split}
\mathrm{RMSE}(\mathbf{x}_k) < T \ \Rightarrow\ \norma{\mathbf{R}-(\mathbf{A}-z(\omega)\mathbf{I})\mathbf{x}_k}_2 < \sqrt{N}\cdot T,
\end{split}
\end{equation}

We can now investigate the accuracy of the iterative solution by adopting different RMSE thresholds. As a precision measure, we consider the longitudinal absorption cross section $\sigma_X$ calculated for a set of GDs (GD20, GD26, GD32, GD36 in \cref{tab:gd-data}) applying an electric field with a polarization vector laying on the molecular plane. The linear system has been solved with both GMRES and a direct procedure, i.e. an LU factorization of the coefficient matrix.\cite{golub2013matrix} For each selected RMSE value, the $\sigma_X$ relative error between GMRES and LU factorization averaged over all the considered frequencies (0.0 eV to 2.0 eV, with a step of 0.1 eV) has been calculated, and the results are graphically depicted in \cref{fig:thresh-gd}. An approximate upper bound of the intrinsic precision associated with the factorization algorithm is also plotted (see section S1 in the SM).

It can be seen that the accuracy of the iterative solution for the different systems is almost constant for a specific RMSE value. In particular, by imposing RMSE$\leq 10^{-4}$, the correct order of magnitude obtained by using the LU solution can be recovered by the iterative procedure, i.e. the relative error is $\leq 10^{-1}$. To further demonstrate that the RMSE criterion is effectively size-independent, we performed the same analysis discussed above also for the CNT case (see Fig. S6 in the SM).

\begin{figure}[!ht]
\includegraphics[width=3.4in]{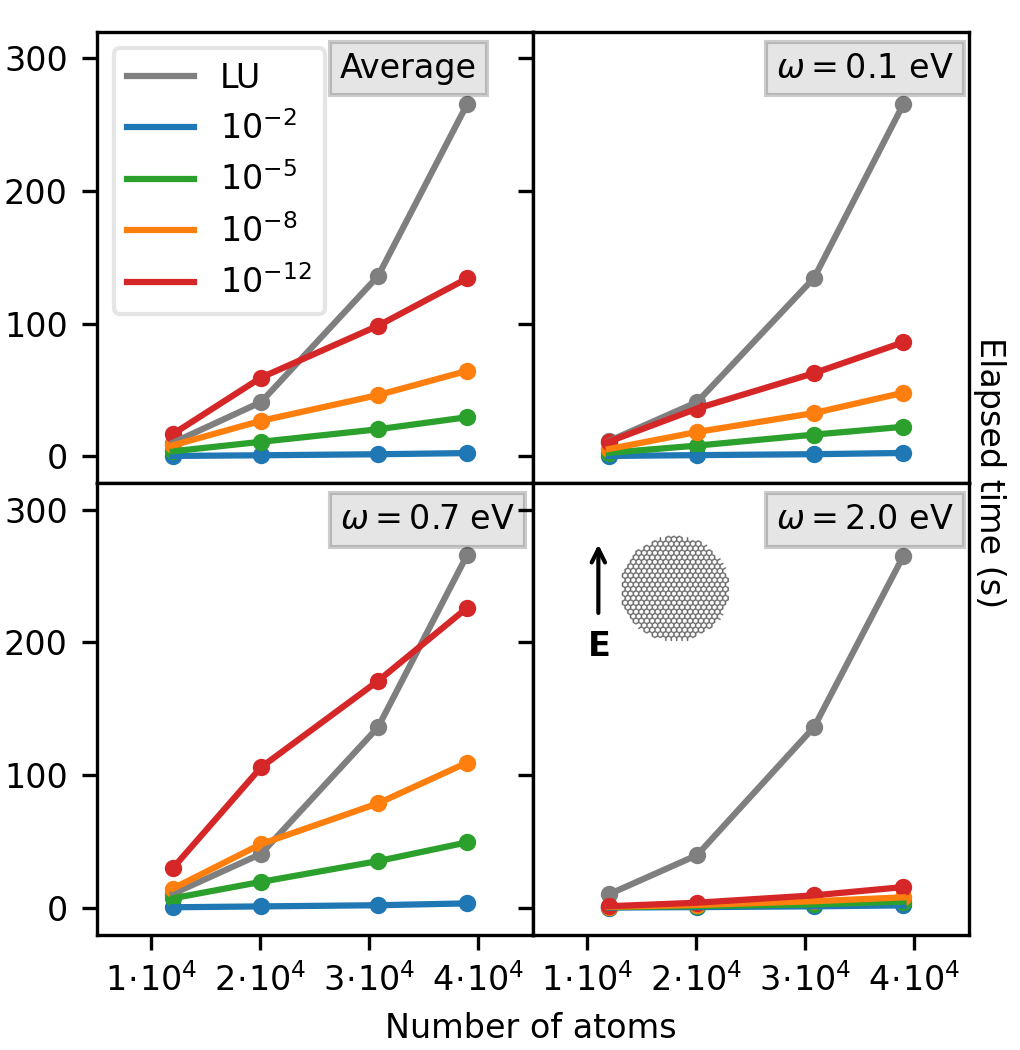}
\caption{Computational time required to solve the \wfq linear system as a function of the number of atoms in GD structures. Average (top, left), $\omega = 0.1$ eV (top, right), $\omega = 0.7$ eV (bottom, left) and $\omega = 2.0$ eV (bottom, right).} 
\label{fig:time}
\end{figure}

We finally move to discuss the computational time required by the different choices of the RMSE value. In particular, such an analysis has been performed on the four GDs reported in \Cref{tab:gd-data}. The  computational time required to solve the \wfq linear system for 20 frequencies in the range between 0.0 and 2.0 eV with a step of 0.1 eV are given in \cref{fig:time} (raw data can be found in Tabs. S2-S5 in the SM). Both direct solution (LU factorization, see section S2 in the SM) and the full GMRES algorithm with different choices of RMSE have been considered. Notice that we are showing the computer time required by the solution of the \wfq linear system. This means that the coefficient matrix $\mathbf{A}$ (\cref{eq:a-matrix-def}) and right-hand side (\cref{eq:r-vector-def}) construction are not taken into account, in order to allow for a direct comparison with the factorization algorithm. All calculations have been performed on a Xeon Gold 5120 (56 cores, 2.2 GHz) cluster node equipped with 256 GB RAM.

In \cref{fig:time}, the average computational time is reported (top, left panel) together with the time required to solve the \wfq linear system for three selected frequencies, i.e. 0.1 (lower bound), 0.7 and 2.0 eV (upper bound). $\omega =0.7$ eV has been chosen because it corresponds to the maximum number of iterations required by full GMRES to converge (see also \cref{fig:disk-total}). 
\Cref{fig:time} clearly shows that the direct solution through LU factorization and the GMRES iterative procedure intrinsically differ in terms of the scaling with the dimension of the linear system (i.e. the number of atoms). In fact, the former has a complexity of $O(N^3)$ while the latter scales as $O(N^2)$. Such a difference is highlighted by the computational times in \cref{fig:time}: the increase of the computational time for the direct solution (gray line) is larger with respect to the iterative procedure when the number of atoms increases, independently of the RMSE value exploited in the GMRES algorithm. Also, we note that, differently from the LU-based algorithm, the computational time required by the iterative method is not constant across the frequency range. This can be explained by the fact that the number of iterations required to converge to the solution depends on the external frequency (see also \cref{fig:disk-total}). Finally, we remark that even at 0.7 eV GMRES is more efficient than the inversion algorithm. In particular, a good compromise between accuracy and computational efficiency can be reached by using RMSE$=10^{-5}$, for which the computational time of the inversion solution can be reduced by a factor of 10 for the largest studied structure. 

The computational time analysis has been performed also for the QMR algorithm (see section S2 and Tabs. S6-S9 in the SM). From an inspection of the numerical results, it emerges that, for a given RMSE, QMR requires roughly twice the computational time than GMRES. In fact, as it has been shown above, GMRES and QMR need a similar number of iterations to converge; however, for each iteration GMRES performs a single matrix-vector product, while QMR computes two matrix-vector products (one with the coefficient matrix, and one with the $\mathbf{D}$ matrix defined in \cref{eq:d-matrix}).
Therefore, although the computational and memory costs of GMRES are not fixed during the iterative procedure, it overperforms QMR because of the lower number of matrix-vector products that are needed to build the Krylov subspace.

\subsection{Large systems}

To finally demonstrate the robustness of the developed iterative methods to solve the \wfq linear system, we investigated the plasmonic response of real-size systems, composed of roughly one million atoms. When dealing with large-sized structures, two main issues arise. From the theoretical point of view, the quasi-static approximation on which \wfq equations are based could be no longer valid.
From the technical point of view, when the number of atoms increases the storing of the \wfq matrix in physical memory can rapidly become unfeasible. Such a problem can be handled by adopting a matrix-free version of the GMRES algorithm, where the $\mathbf{A}$ matrix in \cref{eq:wfq-a-matrix} is not explicitly built. In fact, the iterative algorithm only requires to calculate the matrix-vector product $\mathbf{Ax}$, which can be performed on-the-fly during the execution of the program. This means that at each iterative step $k$, the new Krylov basis vector is obtained as:
\begin{equation}
(\mathbf{x}_k)_i = \sum_{j=1}^N(\mathbf{A} - z(\omega)\mathbf{I})_{ij} (\mathbf{x}_{k-1})_{j},
\end{equation}
where the element $(i,j)$ of the matrix $\mathbf{A}-z(\omega)\mathbf{I}$ is calculated when required by the algorithm. On the other hand, each matrix element has to be calculated from scratch, therefore the on-the-fly version of GMRES would require larger computational time, without affecting the number of iterations. Nevertheless, the on-the-fly matrix-vector product can be efficiently calculated in a parallel environment and memory requirements are negligible with respect to the standard GMRES procedure, because only the iterative vectors should be kept in memory during the solution procedure.


To showcase the performance of GMRES when applied to large systems, we have selected three structures composed by roughly 1 million atoms: a carbon nanotube --CNT1M--, a graphene disk --GD1M-- and a sodium nanorod --NR1M-- (see Tab. S10 given in the SM, for geometrical parameters). The latter is genuinely different from the other two structures and has been selected to further demonstrate the reliability of the method to study the optical properties of metal nanostructures.

For each of the constructed structures, the longitudinal absorption cross sections and the NI have been calculated. We have set RMSE to $10^{-5}$, which has been proved in the previous section to be a good compromise between accuracy and the computational cost. 

\begin{figure}[!h]
\includegraphics[width=3.4in]{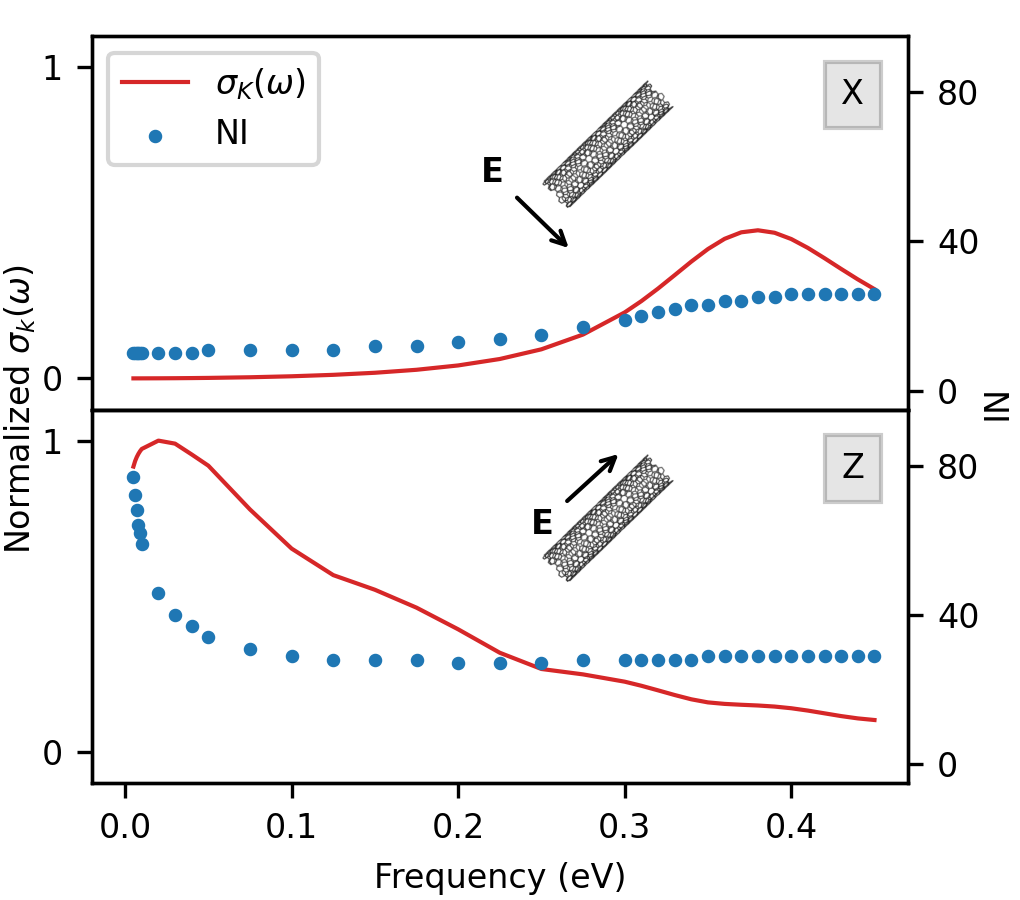}
\caption{CNT1M $\sigma_k(\omega)$ (solid red line) and NI (blue dots). Both longitudinal ($k=Z$, bottom) and transversal ($k=X$, top) polarizations of the external field are reported. GMRES algorithm (RMSE = 10$^{-5}$).}
\label{fig:cnt-big}
\end{figure}

\textbf{CNT1M.} The calculations on CNT1M have been performed by applying an external field along the transversal and longitudinal directions, at 35 different frequencies in the range between 0.0 and 0.45 eV, by setting $\tau=170$ a.u. and $\varepsilon_F = 1.03$ eV. The longitudinal absorption cross sections and the NI are reported in \cref{fig:cnt-big}. The transversal PRF is placed at about 0.38 eV, which is close to the value for CNT4, which has the same diameter (see \cref{tab:cnt-data}). The longitudinal PRF is instead placed at 0.02 eV, which is smaller than the value for CNT300, which has a length of 300 nm. This is not surprising because the PRF is red-shifted as the length of the system increases. 
The required number of iterations as a function of the external field frequency is reported in \cref{fig:cnt-big}. We note that the maximum number of iterations is 80, which is lower than what we have obtained for CNT300 (see \cref{subsec:geometry}). This is due to the larger convergence threshold chosen for the iterative procedure; overall, a mean value of about 30 iterations is sufficient to reach the convergence. Since the number of iterations is modest, restarted GMRES has not been considered for such large systems. Moving on to discuss the computational time, our implementation permits to calculate about 4 matrix-vector product per hour, thus resulting in a total time of about 487 hours. 

\begin{figure}[!h]
\includegraphics[width=3.4in]{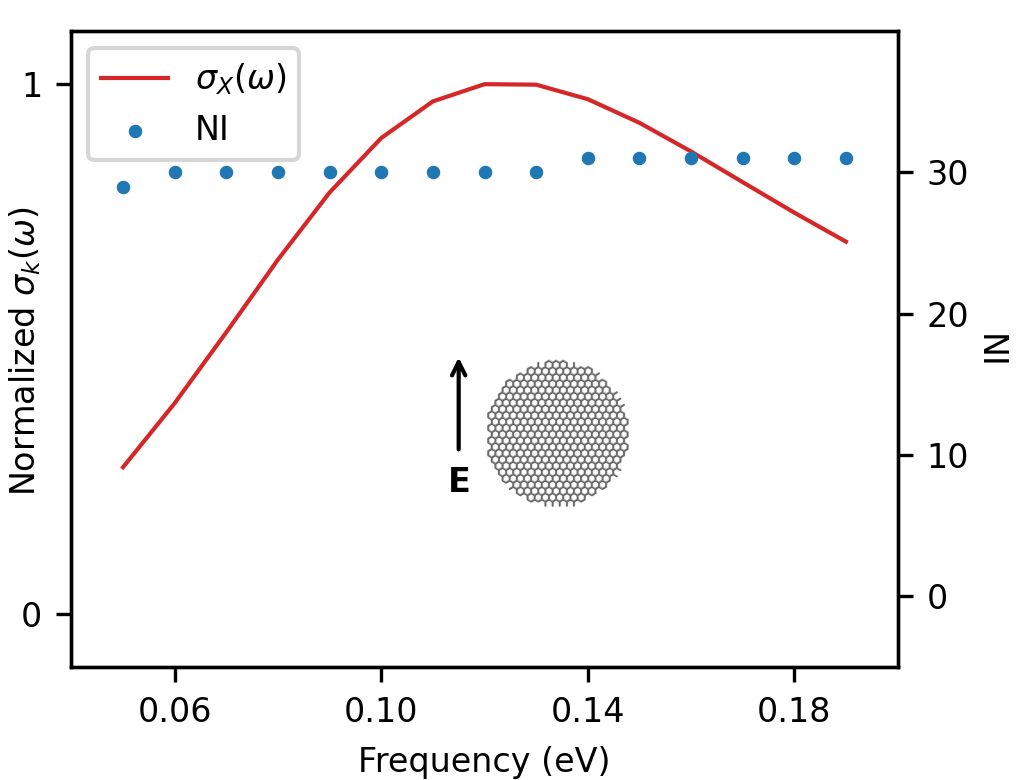}
\caption{GD1M $\sigma_X(\omega)$ (solid red line) and NI (blue dots). GMRES algorithm (RMSE = 10$^{-5}$).}
\label{fig:gd-big}
\end{figure}

\textbf{GD1M.} The \wfq linear system has been solved for GD1M with an external field polarization vector laying on the GD plane ($X$), at 15 different frequencies in the range between 0.05 and 0.19 eV with a constant step of 0.01 eV. We set $\tau=170$ a.u. and $\varepsilon_F = 1.84$ eV. The computed $\sigma_X$ and NI are reported in \cref{fig:gd-big}. The transversal dipolar PRF for this system lays at 0.12 eV, which is smaller than the values for the smallest GDs studied in the previous sections. The required number of iterations to reach convergence is about 30 for each external field frequency, i.e. smaller than what is required by GDs described in \cref{subsec:geometry}. As it has been stated for CNT, this is mainly due to the setting of a larger RMSE threshold.

\begin{figure}[!h]
\includegraphics[width=3.4in]{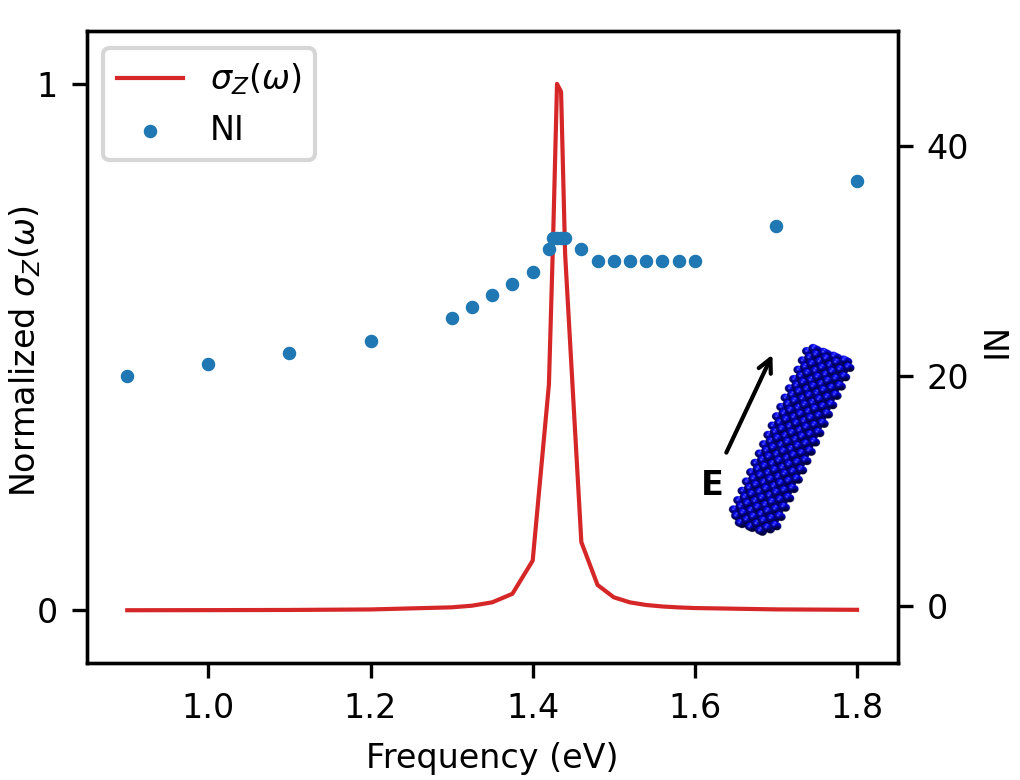}
\caption{NR1M longitudinal $\sigma_Z(\omega)$ (solid red line) and NI (blue dots). GMRES algorithm (RMSE = 10$^{-5}$).}
\label{fig:nanorod-big}
\end{figure}

\textbf{NR1M.} Finally, we have calculated the plasmonic response of the NR1M system. This is a 3D nanostructure, therefore a general expression of the $\mathbf{K}^\mathrm{tot}$ matrix in terms of the 3D electron density $n_0$ (see \cref{eq:k-tot-matrix}) needs to be exploited. All calculations have been performed with \wfq parameters for sodium reported in a previous work. \cite{giovannini2019classical} The linear system in \cref{eq:wfq-a-matrix} has been solved for 24 frequencies in the range between 0.9 and 1.8 eV (unevenly distributed) with an external field aligned along the longitudinal ($Z$) direction. As for CNT1M and GD1M, the RMSE threshold was set to  $10^{-5}$. The computed $\sigma_Z$ and the corresponding NI are reported in \cref{fig:nanorod-big}.

The longitudinal PRF is placed at about 1.43 eV, which is blue shifted with respect to the longitudinal PRF of CNT1M, due to the fact that the electronic properties of the two materials are different. As for the previously studied carbon-based system, the value of the longitudinal PRF is red-shifted with respect to smaller sodium nanorods reported in a previous work.\cite{giovannini2019classical} This is once again due to the lighting rod effect discussed above. Similarly to previous cases, a mean value of about 30 iterations is sufficient to reach convergence.

\section{Summary and Conclusions}

In this paper, we have discussed how to substantially increase the applicability of classical, fully atomistic approaches to the calculation of the plasmonic properties of real-size nanostructures (carbon nanotubes, graphene-based materials and metal nanoparticles). State-of-the-art numerical iterative Krylov-based techniques, GMRES and QMR, have been specified to the recently developed fully atomistic \wfq model. For the studied structures, we have highlighted the dependence of the convergence rate on physical properties and parameters entering the definition of the model. It is worth remarking that a correlation between local maxima of the number of iterations and plasmon resonance frequencies is observed, showing that the convergence rate is particularly slowed down for high-order plasmons. Finally, on-the-fly implementation of GMRES has allowed the calculation of plasmon resonances of very large realistic nanostructures, composed of roughly one million atoms, which are not affordable by direct algorithms. We have also demonstrated the reliability of our implementation by showing that the required number of iterations to correctly describe the plasmon response of such large structures, is moderate.

The implemented iterative procedures are characterized by three main bottlenecks. On one hand, the number of matrix-vector products needed to build approximate solutions to the linear system is associated with a computational complexity of $O(N^2)$. Linear scaling in matrix-vector products might be achieved through the fast multipole method (FMM)\cite{rokhlin1985rapid,engheta1985fast,cipra2000best,lipparini2019general}, a numerical technique that can be adopted to build an approximation to the long-range electrostatic forces, which has already been applied to plasmonic substrates.\cite{payton2014hybrid} Such an extension will allow to afford systems even larger than those studied in this work. However, in this case the quasi-static approximation on which \wfq relies, may be no longer valid. Therefore, retardation effects would need to be included in the model, similarly to what has already been proposed for continuum approaches.\cite{waxenegger2015plasmonics,hohenester2014simulating} A second bottleneck in our procedure is the lack of any preconditioning of the \wfq linear system. Different approaches may be exploited to solve dense complex linear systems, and they will be tested in future works for \wfq. \cite{benzi2002preconditioning,alleon1997sparse,chen2005matrix}
Finally, in order to study the plasmonic properties of a given nanostructure along a specific spectral region, the \wfq linear system can be solved independently for each frequency. However, a change in frequency only affects the uniform diagonal shift of the coefficient matrix in \cref{eq:wfq-a-matrix}. Therefore, the shift-invariance property of the Krylov subspaces may be exploited by resorting to the so-called subspace recycling techniques.\cite{sun2018block,soodhalter2020survey}

To conclude, another valuable application of \wfq is its potential applicability to theoretically describe Surface-Enhanced spectroscopies, either based on graphene-based substrates or metal nanoparticles.\cite{langer2019present,ling2015lighting} To this end, \wfq needs to be coupled with a quantum Hamiltonian describing the adsorbed moiety, in a QM/MM fashion.\cite{morton2010discrete,morton2011discrete,payton2012discrete,payton2014hybrid,chen2015atomistic} Such a further development is on-going in our group and will be the topic of future communications.

\section{Acknowledgment}

This work has received funding from the European Research Council (ERC) under the European Union’s Horizon 2020 research and innovation programme (grant agreement No. 818064). 

\medskip

\textbf{Data Availability Statement}
The data that support the findings of this study are available from the corresponding author upon reasonable request.


\bibliography{biblio}

\end{document}